\documentclass{achemso}

\usepackage{amsmath}
\usepackage{amssymb}
\usepackage{dcolumn}
\usepackage{graphicx}
\usepackage[version=3]{mhchem} % Formula subscripts using \ce{}

\newcommand{\be}{\begin{equation}}
\newcommand{\ee}{\end{equation}}
\newcommand{\bea}{\begin{eqnarray}}
\newcommand{\eea}{\end{eqnarray}}
\newcommand{\bm}[1]{\mathbf{#1}}

\def \ve{{\varepsilon}}

\def \cE{{\cal E}}
\def \cC{{\cal C}}
\def \cQ{{\cal Q}}
\def \cW{{\cal W}}

\title{Electronic Cooling via Interlayer Coulomb Coupling in Multilayer Epitaxial Graphene}

\author{Momchil T. Mihnev}
\affiliation{Department of Electrical Engineering and Computer Science, University of Michigan, Ann Arbor, MI 48109, USA\\}
\alsoaffiliation{Center for Ultrafast Optical Science, University of Michigan, Ann Arbor, MI 48109, USA\\}
\altaffiliation{These authors contributed equally}

\author{John R. Tolsma}
\affiliation{Department of Physics, The University of Texas at Austin, Austin Texas 78712, USA\\}
\altaffiliation{These authors contributed equally}

\author{Charles J. Divin}
\affiliation{Department of Electrical Engineering and Computer Science, University of Michigan, Ann Arbor, MI 48109, USA\\}
\alsoaffiliation{Center for Ultrafast Optical Science, University of Michigan, Ann Arbor, MI 48109, USA\\}

\author{Dong Sun}
\affiliation{Center for Ultrafast Optical Science, University of Michigan, Ann Arbor, MI 48109, USA\\}
\alsoaffiliation{International Center for Quantum Materials, School of Physics, Peking University, Beijing 100871, People’s Republic of China\\}

\author{Reza Asgari}
\affiliation{School of Physics, Institute for Research in Fundamental Sciences (IPM), Tehran 19395-5531, Iran\\}

\author{Marco Polini}
\affiliation{NEST, Istituto Nanoscienze-CNR and Scuola Normale Superiore, I-56126 Pisa, Italy\\}
\alsoaffiliation{Istituto Italiano di Tecnologia, Graphene Labs, Via Morego 30, I-16163 Genova, Italy\\}

\author{Claire Berger}
\affiliation{School of Physics, Georgia Institute of Technology, Atlanta, GA 30332, USA\\}
\alsoaffiliation{Institut Neel, CNRS UJF-INP, 38042 Cedex 6, Grenoble, France\\}

\author{Walt A. de Heer}
\affiliation{School of Physics, Georgia Institute of Technology, Atlanta, GA 30332, USA\\}
\alsoaffiliation{King Abdulaziz University, Jeddah 22254, Saudi Arabia\\}

\author{Allan H. MacDonald}
\affiliation{Department of Physics, The University of Texas at Austin, Austin Texas 78712, USA\\}

\author{Theodore B. Norris}
\affiliation{Department of Electrical Engineering and Computer Science, University of Michigan, Ann Arbor, MI 48109, USA\\}
\alsoaffiliation{Center for Ultrafast Optical Science, University of Michigan, Ann Arbor, MI 48109, USA\\}
\email{tnorris@umich.edu}
\phone{+1-734-764-9269}
\fax{+1-734-763-4876}

\begin{document}
%%%%%%%%%%%%%%%%%%%%%%%%%%%%%%%%%%%%%%%%%%%%%%%%%%%%%%%%%%%%%%%%%%%%%
%% Main text
%%%%%%%%%%%%%%%%%%%%%%%%%%%%%%%%%%%%%%%%%%%%%%%%%%%%%%%%%%%%%%%%%%%%%

%% Introductory paragraph / Abstract
{\bf In van der Waals bonded or rotationally disordered multilayer stacks of two-dimensional (2D) materials, the electronic states remain tightly confined within individual 2D layers. As a result, electron-phonon interactions occur primarily within layers and interlayer electrical conductivities are low. In addition, strong covalent in-plane intralayer bonding combined with weak van der Waals interlayer bonding results in weak phonon-mediated thermal coupling between the layers. 
We demonstrate here, however, that Coulomb interactions between electrons in different layers of multilayer epitaxial graphene provide an important mechanism for interlayer thermal transport even though all electronic states are strongly confined within individual 2D layers. This effect is manifested in the relaxation dynamics of hot carriers in ultrafast time-resolved terahertz spectroscopy. We develop a theory of interlayer Coulomb coupling containing no free parameters that accounts for the experimentally observed trends in hot-carrier dynamics as temperature and the number of layers is varied.} 

%% Introduction
\section{Introduction}
The dynamics of electrons in atomic-layer 2D electron systems like graphene is a subject of considerable current interest, partly because of its relevance to a wide variety of potential electronic and optoelectronic device applications. Many proposed and prototype devices employ stacks composed of many 2D electronic material layers. Examples of multilayer systems include multilayer epitaxial graphene (MEG) \cite{deHeerScience2006, deHeerPNAS2011}, van der Waals bonded layered sheets \cite{GeimNature2013, BritnellScience2013}, transition metal dichalcogenides \cite{StranoNatNano2012} and others \cite{ZhongNatNano2014}. In these structures the interactions between electrons in different layers in the stack becomes a subject of key importance. One important property is that phonon-mediated interlayer thermal coupling is weak relative to that in bulk 3D materials. 
In the example of MEG, rotational stacking arrangements decouple electronic states localized in different 2D layers \cite{deHeerPRL2008, deHeerPRL2009}. As a result, phonon-mediated interlayer thermal coupling in MEG is strongly reduced relative to typical bulk behavior in 3D materials. 

The question thus arises as to what other mechanisms can contribute to thermal equilibration between different layers. 
We consider this question here in the context of hot carrier dynamics. If electrons are heated in one layer (e.g. by optical excitation or electrical injection), they will normally cool to the lattice temperature by optical phonon emission at high carrier energies, and by acoustic phonon emission at low carrier energies. For graphene, it is well established that electron cooling by acoustic phonons is very efficient in highly doped layers \cite{BistritzerPRL2009, TsePRB2009, SunPSSC2011}. The situation is quite different, however, for lightly doped or nearly neutral graphene in which a small joint-density-of-states for electronic transitions combines with a
small acoustic phonon energy at typical scattering wavevectors to diminish the acoustic phonon cooling power \cite{BistritzerPRL2009, TsePRB2009}. 
(The cooling power in this limit can, however, be substantially enhanced by disorder, because it relaxes momentum conservation limits \cite{LevitovPRL2012, GrahamNatPhys2013, BetzNatPhys2013} on allowed processes.) 

An interesting case thus arises when a multilayer stack contains both highly doped (HD) and nearly neutral lightly doped (LD) graphene layers. 
This is exactly the situation that occurs in MEG grown on the C-face of SiC \cite{deHeerScience2006, deHeerPNAS2011}, and is likely to be relevant to gated multilayer 2D systems due to interlayer screening \cite{SunPRL2008, SunPRL2010}. If electrons are heated in the multilayer structure, then acoustic-phonon-mediated cooling would result in the rapid buildup of a thermal gradient between the HD and LD layers; the HD layers would quickly approach the lattice temperature, while carriers in the LD layers would remain hot. Eventually, of course, thermal equilibrium would be restored, as thermal energy flows from the LD to the HD layers. 
The HD layers can be a heat sink for the LD layers if there is an effective interlayer energy transfer mechanism.  

In the following, we show that Coulomb scattering between electrons in LD and HD layers of MEG can provide an efficient 
means for interlayer thermal coupling, and provide an alternate mechanism for cooling of hot electrons in the LD layers that acts
in parallel with acoustic-phonon-mediated intralayer cooling. 
This process is illustrated schematically in Figure~\ref{fig:fig1}a. 
We note that interlayer thermal coupling via Coulomb scattering has been considered recently in the context of 2D electron gases in transport devices \cite{CoppersmithPRB2011, PrunnilaNJP2013}. 
We begin by outlining a heuristic analytical model calculation for a pair of graphene layers to establish the magnitude of the effect relative to acoustic-phonon-mediated intralayer cooling. This simple calculation establishes the significance of the effect; we then discuss how hot-carrier cooling in multilayer systems is accessible experimentally by ultrafast time-resolved terahertz (THz) spectroscopy and ultrafast infrared (IR) pump-probe spectroscopy measurements. Following a discussion of the features of the data that point to interlayer energy transfer, we present the details of a theory of interlayer energy transfer via screened Coulomb interactions. The calculated cooling powers imply asymptotic cooling times on the sub-nanosecond scale. We show that the calculated dynamics and trends with lattice temperature and number of epitaxial graphene layers are fully consistent with the experimental results, without the need for any fitting parameters. 

%% Results
\section{Results}

%% Heuristics
\subsection{Interlayer Coulombic energy transfer heuristics}
\begin{figure}
\begin{center}
\includegraphics[width=6.0in]{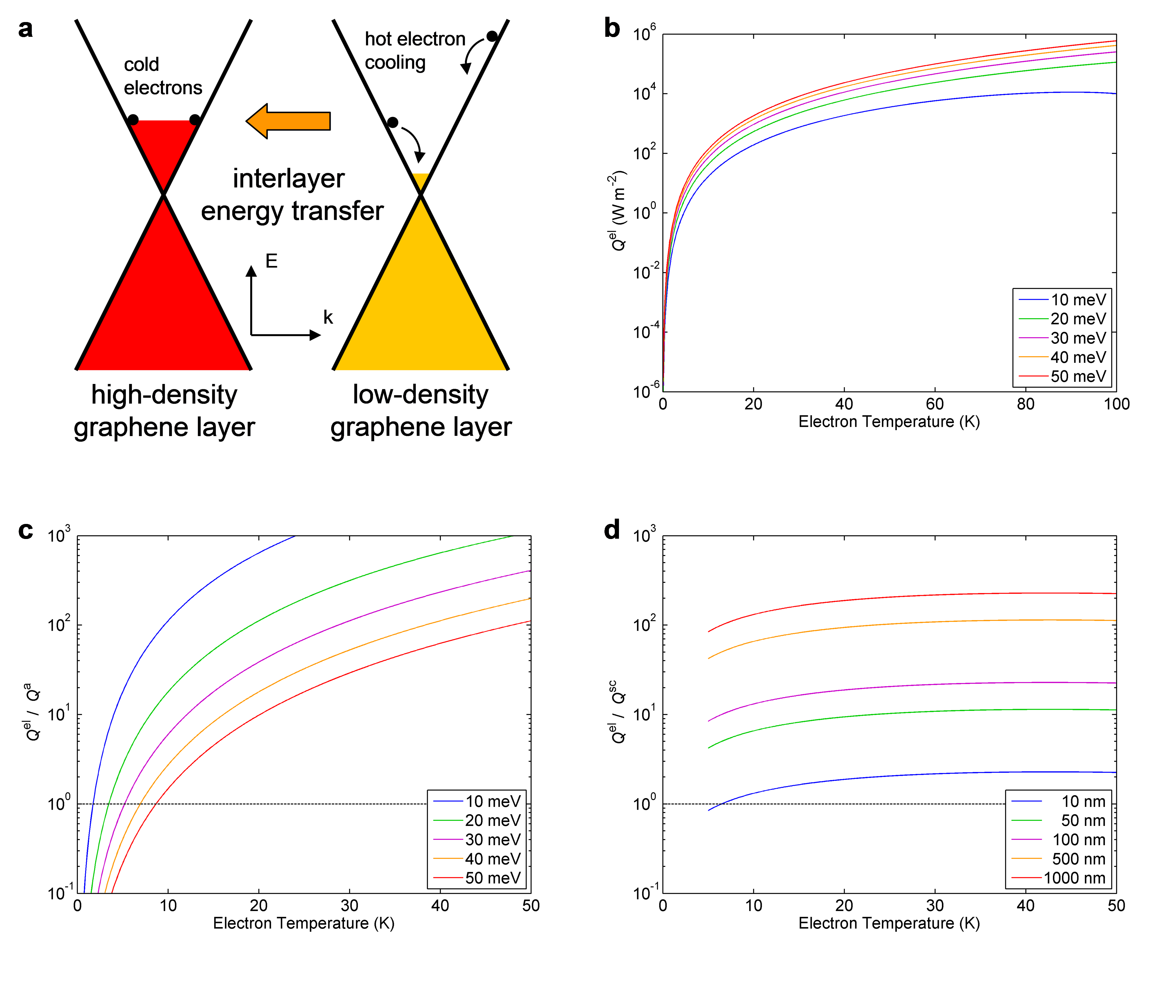}
\caption{{\bf Interlayer Coulombic energy transfer between two graphene layers} 
a, Schematic diagram of interlayer Coulombic energy transfer from a hot LD to a cold HD graphene layer. 
b, Cooling power of interlayer Coulombic energy transfer $\cQ^{\rm{el}}$ as a function of electron temperature for Fermi level $E_{\rm{F,HD}} = 300$ meV in the HD graphene layer and various Fermi levels $E_{\rm{F,LD}}$ in the LD graphene layer. 
c, Ratio of the cooling power of interlayer Coulombic energy transfer $\cQ^{\rm{el}}$ to the cooling power of intralayer acoustic phonon cooling $\cQ^{\rm{a}}$ as a function of electron temperature for $E_{\rm{F,HD}} = 300$ meV and various values of $E_{\rm{F,LD}}$ showing that interlayer Coulombic energy transfer can dominate intralayer acoustic phonon cooling in the LD graphene layer. 
d, Ratio of the cooling power of interlayer Coulombic energy transfer $\cQ^{\rm{el}}$ to the cooling power of intralayer disorder-assisted electron-phonon (supercollision) cooling $\cQ^{\rm{sc}}$ as a function of electron temperature for $E_{\rm{F,HD}} = 300$ meV, $E_{\rm{F,LD}} = 10$ meV (typical for C-face MEG on SiC) and various values of the disorder mean free path in the LD graphene layer showing that interlayer Coulombic energy transfer can dominate intralayer disorder-assisted electron-phonon (supercollision) cooling in the LD graphene layer.}
\label{fig:fig1}
\end{center}
\end{figure}

We first ask whether or not interlayer Coulomb coupling can potentially dominate over acoustic phonon cooling \cite{BistritzerPRL2009} and disorder-assisted electron-phonon (supercollision) cooling \cite{LevitovPRL2012} in multilayer graphene samples. 
A simple comparison of the cooling powers of the different mechanisms suggests that the answer is yes. 
The low-temperature cooling power $\cQ^{\rm {el}}$ of interlayer Coulombic energy transfer between a hot LD and a cold HD graphene layer is (see Supplementary Notes 2-3): 
\begin{equation}\label{eq:eratelo}
\cQ^{\rm {el}} = \frac{\pi^2 k_{\rm B}^4 \nu_{\rm {LD}}}{15 \hbar^3 v_{\rm F}^2 \nu_{\rm {HD}}}T^4 \ln \left(\frac{E_{\rm {F,LD}}}{k_{\rm B}T} \right), 
\end{equation}
\noindent
where $\nu_i = 2E_{{\rm F},i}/(\pi \hbar^2 v_{\rm F}^2)$ is the density of states at the Fermi level. 
Equation~\ref{eq:eratelo} is plotted in Figure~\ref{fig:fig1}b as a function of electron temperature for 
Fermi level $E_{\rm {F,HD}} = 300$ meV in the HD layer and various Fermi levels $E_{\rm {F,LD}}$ in the LD layer; even at very low carrier density in the LD layer, the cooling power is quite substantial. 
We can compare Equation~\ref{eq:eratelo} with the cooling powers of both acoustic phonon cooling $\cQ^{\rm {a}} \propto T$ and disorder-assisted electron-phonon (supercollision) cooling $\cQ^{\rm {sc}} \propto T^{3}$ of an isolated LD layer. 
The ratio of cooling powers $\cQ^{\rm {el}} / \cQ^{\rm {a}}$ is plotted in Figure~\ref{fig:fig1}c as a function of electron temperature for $E_{\rm {F,HD}} = 300$ meV and various values of $E_{\rm {F,LD}}$. 
As it can be expected, acoustic phonon cooling is very inefficient in graphene with very low carrier density. 
The ratio of cooling powers $\cQ^{\rm {el}} / \cQ^{\rm {sc}}$ is also plotted in Figure~\ref{fig:fig1}d as a function of electron temperature (above the Bloch-Gr{\"u}neisen temperature $T_{{\rm BG}} \approx 5$ K) for $E_{\rm {F,HD}} = 300$ meV, $E_{\rm {F,LD}} = 10$ meV (typical for C-face MEG on SiC) and various values of the low-density-layer disorder mean free path. 
It is apparent that for high quality graphene such as C-face MEG on SiC \cite{deHeerNature2014}, interlayer Coulombic energy transfer can dominate for a wide range of electron temperatures and sample characteristics. 

%% MEG samples
\subsection{Multilayer epitaxial graphene}
In the main body of this paper, we investigate the physics of interlayer Coulomb coupling in MEG, grown on the C-face of single-crystal 4H-SiC($000\overline{1}$) substrates by thermal decomposition of Si atoms \cite{deHeerScience2006, deHeerPNAS2011}. An important feature of this material is that it largely preserves distinct graphene-like electronic properties because of unique rotational stacking, which suppresses hybridization between low energy electronic states localized in neighboring planes of carbon atoms \cite{deHeerPRL2008, deHeerPRL2009}. MEG is doped by electron transfer from the interface with the supporting SiC substrate and the induced n-type carrier-density profile falls off rapidly with layer moving away from the substrate (see the inset of Figure~\ref{fig:fig2}a). We will refer to the few layers close to the SiC substrate, which have large carrier densities of $n_{\rm {HD}} \gtrsim 10^{12}$cm$^{-2}$ as determined from high-resolution angle-resolved photoemission spectroscopy (ARPES), scanning tunneling spectroscopy (STS), electronic transport, and ultrafast optical spectroscopy measurements \cite{deHeerScience2006, deHeerPNAS2011, ConradJPDAP2012, SunPRL2008, SunPRL2010}, as high-density (HD) layers, and to those further away, whose carrier densities drop quickly \cite{Hongki} to the range of $n_{\rm {LD}} \lesssim 10^{10}$cm$^{-2}$ as determined from STS, electronic transport, and magneto-optical spectroscopy measurements \cite{deHeerScience2006, PotemskiPRL2011}, as low-density (LD) layers. The formation of local spatial charge inhomogeneities due to small amounts of disorder, impurities, or surface corrugation of the SiC substrate could explain the non-zero carrier density measured in the top layers of MEG \cite{StroscioNatPhys2010, deHeerAPL2009, GaskillAPL2011}. 

\subsection{Electronic cooling in multilayer epitaxial graphene}
When a MEG sample is illuminated with a short optical pulse, electrons are excited to high energies, leaving behind unoccupied states or holes. Due to strong intralayer carrier-carrier scattering, these hot carriers thermalize with the background of cold carriers within $\sim 50$ fs \cite{BreusingPRL2009, BreusingPRB2011, BridaNatComm2013, GilbertsonJPCLett2012, CavalleriNatMat2013}, forming two separate non-equilibrium Fermi-Dirac distributions for electrons and for holes. The electron and the hole quasi-Fermi levels subsequently merge within $\sim 100-200$ fs \cite{GilbertsonJPCLett2012, CavalleriNatMat2013}, establishing a uniform electron liquid within each layer $i$ characterized by an elevated electron temperature, $T_i$. As the electron liquid cools, each layer's electron temperature approaches the equilibrium lattice temperature, $T_{\rm L}$. It is generally accepted that initial fast cooling occurs in the first few picoseconds via the emission of energetic optical phonons, and that this process becomes increasingly inefficient as the electron energy falls below the relatively high optical phonon energy ($\hbar\omega_{\rm {op}} \approx 200$ meV \cite{deHeerAPL2008}). In the final stage of relaxation, low energy electronic cooling in an isolated layer would proceed by the much slower emission of acoustic phonons. Recent findings suggest that that the low energy phonon spectra of multilayer graphene systems are sensitive to the pattern of relative orientations \cite{NikaAPL2014}. Although this property will transfer a corresponding sensitivity to phonon cooling powers, relative orientations will not influence the interlayer Coulombic energy transfer mechanism explored here. Prior theoretical work has found that the rate of acoustic phonon cooling in disorder-free single-layer graphene is very strongly dependent on the carrier density, with cooling times ranging from tens of picoseconds for doping densities of $\sim 10^{13}$cm$^{-2}$ to tens of nanoseconds for doping densities of $\sim 10^{10}$cm$^{-2}$ \cite{BistritzerPRL2009, TsePRB2009}. The strong carrier density dependence of the acoustic phonon emission implies that the LD and HD layers of MEG will exhibit very differing cooling rates following ultrafast optical excitation, leading to the buildup of a thermal gradient, which triggers an interlayer energy transfer. 

We have applied two different experimental techniques to probe the dynamics of the interlayer energy transfer in MEG. The first is ultrafast time-resolved THz spectroscopy, which is a powerful tool for investigating the real time relaxation dynamics of photoexcited carriers, because it is sensitive to both the number of carriers and their distribution in energy \cite{SchmuttenmaerCR2004, BaxterAC2011}. Due to the large number of layers in our MEG samples and the rapid decrease of carrier density with layer number, the measured differential THz transmission signal is dominated by the dynamic THz response of the many LD layers and reveals the cooling of the LD layers due to their coupling to the HD layers. The second is ultrafast degenerate IR pump-probe spectroscopy, in which we optically inject hot carriers selectively into the LD layers and then observe directly the transfer of heat to the electron liquid in the most highly doped HD layer \cite{SunPRL2008, SunPRL2010}. 

%% Experiment
% THz Experiment
\subsection{Ultrafast time-resolved THz spectroscopy}
\begin{figure}
\begin{center}
\includegraphics[width=3.25in]{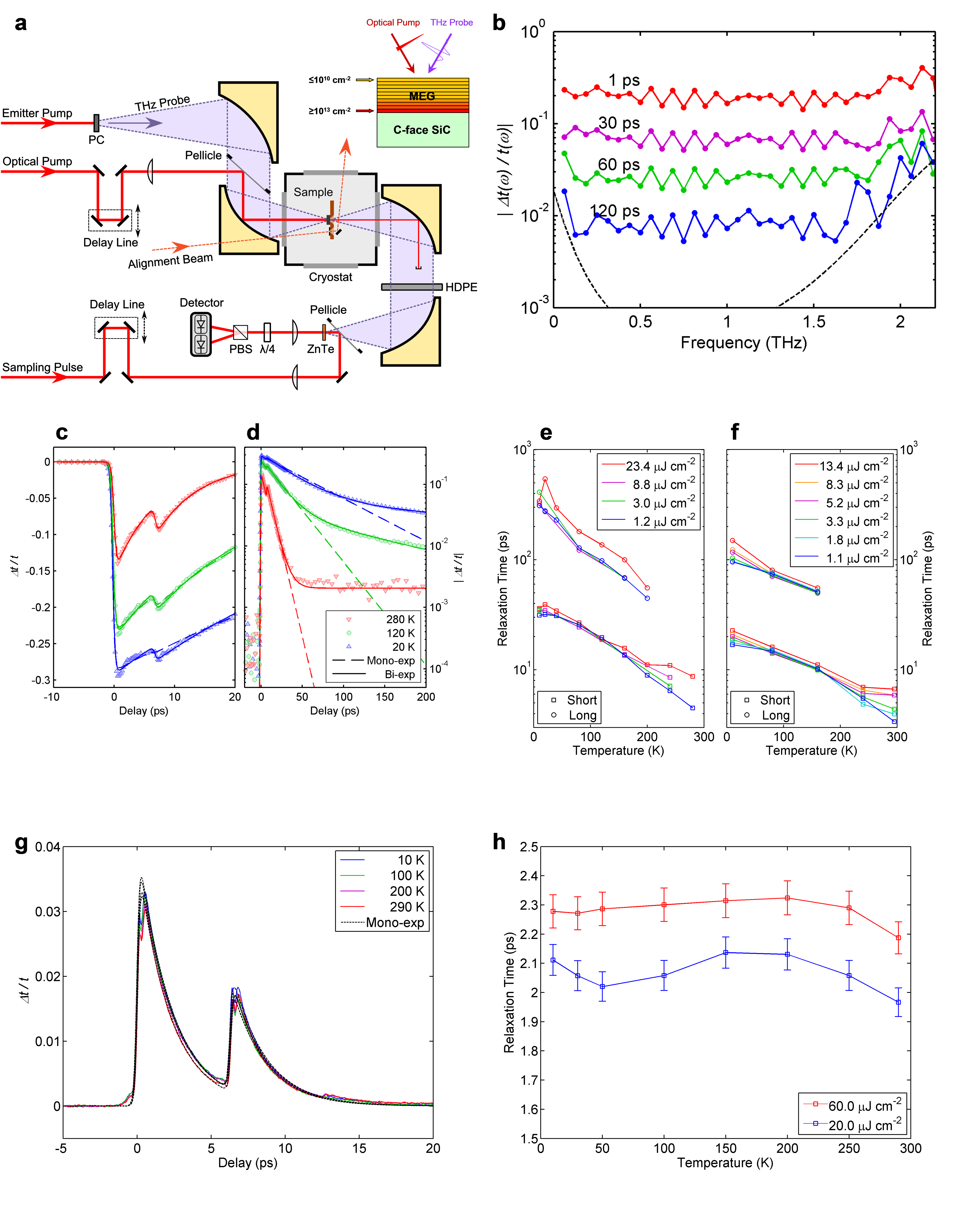}
\caption{{\bf Ultrafast time-resolved THz spectroscopy on MEG} 
a, Schematic diagram of the ultrafast time-resolved THz spectroscopy set-up. (Inset: Schematic diagram of a MEG sample with a gradient doping density profile.) 
b, Normalized differential THz transmission spectra $\Delta t(\omega)/t(\omega)$ recorded at a pump fluence of $0.87$ $\mu$J cm$^{-2}$ and a substrate temperature of $40$ K for a few different pump-probe delays for a MEG sample with $\sim 63$ layers. The black dashed line indicates the experimental noise level. The THz spectra are remarkably dispersionless in the detectable frequency range under all experimental conditions. 
c-d, Linear (c) and logarithmic (d) plots of normalized differential THz transmission at the peak of the THz probe pulse $\Delta t/t$ as a function of pump-probe delay recorded at a pump fluence of $23.4$ $\mu$J cm$^{-2}$ for a few different substrate temperatures for a MEG sample with $\sim 63$ layers. The THz carrier dynamics evolve from a faster mono-exponential relaxation at room temperature to a slower bi-exponential relaxation at cryogenic temperatures. Subfigures (c) and (d) share the same legend. 
e-f, Short and long relaxation times as a function of substrate temperature for a few different pump fluences for a MEG sample with $\sim 63$ (e) and $\sim 35$ (f) layers. The values are extracted from phenomenological fits to normalized differential THz transmission $\Delta t/t$. The long relaxation times increase with the number of epitaxial graphene layers, which indicates the presence of interlayer interaction in MEG with HD and LD layers. 
g, Normalized differential THz transmission at the peak of the THz probe pulse $\Delta t/t$ as a function of pump-probe delay recorded at a pump fluence of $60.0$ $\mu$J cm$^{-2}$ for a few different substrate temperatures for a MEG sample with $\sim 3$ layers. The THz carrier dynamics follow a fast mono-exponential relaxation at all temperatures. 
h, Relaxation times as a function of substrate temperature for a few different pump fluences for a MEG sample with $\sim 3$ layers. The relaxation times are completely independent of the substrate temperature, because there is practically very little or no interlayer energy transfer in MEG with all HD layers. 
The experimental error in all relaxation times is due primarily to long-term drift of the optomechanical components and the ultrafast Ti:Sapphire laser system, and is estimated not to exceed $\sim 5 \%$.}
\label{fig:fig2}
\end{center}
\end{figure}

We first consider the ultrafast time-resolved THz spectroscopy experiments. We report measurements on a series of MEG samples ranging from $\sim 3$ to $\sim 63$ layers in an ultrafast optical-pump THz-probe set-up \cite{SchmuttenmaerCR2004, BaxterAC2011} as illustrated schematically in Figure~\ref{fig:fig2}a. Our laser system consists of a Ti:Sapphire oscillator (Mira $900$-F, Coherent) followed by a Ti:Sapphire regenerative amplifier (RegA $9050$, Coherent) and produces ultrafast optical pulses with a center wavelength of $800$ nm, a pulse width of $\sim 60$ fs and a repetition rate of $250$ kHz. A portion of the laser beam is quasi-collimated at the sample position with an intensity spot size diameter of $\sim 1600$ $\mu$m, and optically injects hot carriers in the MEG samples. A second portion of the laser beam is used to generate a single-cycle THz pulse in a low temperature grown GaAs photoconductive emitter (Tera-SED $3/4$, Gigaoptics) \cite{GrischkowskyAPL1991, TaniAO1997} and the emitted broadband THz radiation is focused on the MEG sample with an intensity spot size diameter of $\sim 500$ $\mu$m to probe the dynamic THz response. The transmitted portion of the THz probe is detected by using time-domain electro-optic sampling in a $1$ mm thick ZnTe crystal \cite{ZhangAPL1995, ZhangAPL1996, HeinzAPL1996} and a pair of balanced Si photodiodes. The electrical signal is modulated by a mechanical chopper, placed in either the optical pump or the THz probe arm, and recorded by using a conventional lock-in amplifier data acquisition technique. The MEG sample is mounted inside a liquid helium continuous flow cryostat (ST-100, Janis), so that the substrate temperature can be varied from $10$ K to $300$ K. The time delays between the optical pump, the THz probe and the sampling pulse are controlled by two motorized stages. All THz optics is surrounded by an enclosure purged with purified nitrogen gas to minimize water vapor absorption. The detection bandwidth of the system is in the range of $\sim 0.2-2.5$ THz and the temporal resolution of the measurements is limited by the duration of the THz probe pulse to the sub-picosecond timescale. The experimental error is due primarily to long-term drift of the optomechanical components and the ultrafast Ti:Sapphire laser system, and is estimated not to exceed $\sim 5 \%$. 

As a first experimental approach, we measure the differential change in the THz probe pulse transmission through the MEG sample due to photoexcitation. The THz probe field is recorded in the time-domain, and it is later numerically Fourier transformed to obtain the frequency spectrum. Figure~\ref{fig:fig2}b shows the differential THz transmission spectra normalized to the THz transmission without photoexcitation, $\Delta t(\omega)/t(\omega)$, for a few different THz probe delays after the optical pump for a MEG sample with $\sim 63$ layers. From the Tinkham formula for the transmission through a thin conducting film on a transparent substrate \cite{TinkhamPR1956}, the $\Delta t(\omega)/t(\omega)$ signal can be directly related to the photoinduced change in the complex sheet conductivity of MEG, $\Delta \sigma(\omega)$, through the expression: 
\begin{equation}\label{eq:sigma}
\Delta \sigma(\omega) \approx -\frac{n_{\rm {sub}}+n_{\rm {vac}}}{\eta_0} \times \frac{\Delta t(\omega)}{t(\omega)}, 
\end{equation}
\noindent
where $n_{\rm {sub}}$ and $n_{\rm {vac}}$ are the THz refractive indices of the SiC substrate and the environment, respectively, and $\eta_0$ is the impedance of free space. The THz conductivity of MEG is a summation of the THz conductivities of the individual epitaxial graphene layers in the MEG stack, because the layers are electronically decoupled \cite{deHeerPRL2008, deHeerPRL2009}. Additional THz transmission spectra, similar to the ones in Figure~\ref{fig:fig2}b, but for variable substrate temperature and variable pump fluence, are shown in Supplementary Fig. 1-2 and Supplementary Note 1. We note that the normalized differential THz transmission spectra are remarkably dispersionless in the detectable frequency range under all experimental conditions; this justifies the application of a simpler data acquisition scheme in which we record the normalized differential THz transmission only at the peak of the THz probe pulse. 

As a second experimental approach, we keep the delay of the sampling pulse fixed at the peak of the THz probe pulse, and we scan the pump-probe delay to map out the relaxation dynamics of the photoexcited carriers. Since the carrier-carrier scattering time in graphene is much shorter than the temporal duration of the THz probe, we study the relaxation of the THz transmission (or the THz conductivity) change induced by the optical pump in the limit, where it is determined by collective electronic cooling dynamics. Figure~\ref{fig:fig2}c-d show the normalized differential THz transmission at the peak of the THz probe pulse, $\Delta t/t$, as a function of pump-probe delay for variable substrate temperature for the same MEG sample with $\sim 63$ layers. At time zero, the optical pump photoexcites carriers in the MEG sample resulting in an overall increase of the THz conductivity and hence THz absorption, as manifested in a negative differential THz transmission. In graphene with very low doping, the increase of the electron temperature leads primarily to larger electron occupation in the conduction band and a corresponding net increase of the THz conductivity, consistent with our interpretation that the measured dynamic THz response is dominated by the hot carriers in the many LD layers of MEG. \cite{RanaNanoLett2011, HeinzNanoLett2013, WangNanoLett2014} The differential THz transmission reaches its maximum magnitude within $\sim 1$ ps, with the rise time being limited mainly by the temporal duration of the THz probe. The differential THz transmission subsequently recovers as the thermalized hot carriers cool to the substrate temperature with relaxation times ranging from a few picoseconds at room temperature to hundreds of picoseconds at cryogenic temperatures. The secondary decrease in the differential THz transmission at $\sim 7$ ps is due to a round-trip reflection of the optical pump inside the substrate that photoexcites additional carriers. 

We perform phenomenological fits to the experimental data in Figure~\ref{fig:fig2}c-d, and we discover that the differential THz transmission evolves from a faster mono-exponential decay at room temperature to a slower bi-exponential decay at cryogenic temperatures. As we explain below, the slow electronic cooling at low substrate temperatures is controlled by interlayer Coulombic energy transfer between the LD and HD layers. A summary of the extracted carrier relaxation times as a function of substrate temperature for a few different pump fluences is presented in Figure~\ref{fig:fig2}e. We observe that the relaxation times are largely independent of the pump fluence (or the initial electron temperature) except at high substrate temperatures. The slight increase in the relaxation times at the highest pump fluence can be attributed to heating of the HD layers above the substrate temperature, which decreases the rate of the energy transfer from the LD layers. Similarly, the energy transfer between layers becomes less efficient at high substrate temperatures, at which the difference between the electron temperatures in different layers is small. As a consequence, the contribution of the interlayer Coulomb coupling to electronic cooling is diminished at substrate temperatures above $\sim 200$ K as evidenced from the fits. 

We repeat identical experiments and analysis for a second MEG sample with $\sim 35$ layers and the corresponding summary of the extracted carrier relaxation times for the best fits are presented in Figure~\ref{fig:fig2}f. Qualitatively, the THz carrier dynamics for the 35-layer sample mirror those for the 63-layer sample by exhibiting a transition from a faster mono-exponential to a slower bi-exponential decay as the substrate temperature is decreased. Similarly, the relaxation times are independent of the pump fluence except at high substrate temperatures. Further inspection shows that the long relaxation times become up to a few times shorter, when the number of epitaxial graphene layers is nearly halved, which indicates the presence of interlayer interaction. We show below that because of the range dependence of the Coulomb scattering processes, the addition of more LD layers slows their collective electronic cooling via coupling to the HD layers. 

To underscore the profound influence of interlayer energy transfer on the electronic cooling in MEG, we next study the limiting case of MEG with all HD layers. We again perform identical experiments and analysis for a third MEG sample with only $\sim 3$ layers. Figure~\ref{fig:fig2}g shows the normalized differential THz transmission at the peak of the THz probe pulse, $\Delta t/t$, as a function of pump-probe delay for variable substrate temperature. First, we note that the differential THz transmission for MEG with all HD layers is positive, which has been previously phenomenologically attributed to enhanced carrier scattering as the electron temperature is elevated \cite{HeinzNanoLett2013, WangNanoLett2014}. Second, we observe that the THz carrier dynamics are much faster and completely independent of the substrate temperature, because there is practically very little or no interlayer energy transfer. The differential THz transmission is fit very well by a phenomenological mono-exponential decay at all temperatures (again accounting for the substrate reflection of the optical pump) and the summary of the extracted carrier relaxation times as a function of substrate temperature for a few different pump fluences is presented in Figure~\ref{fig:fig2}h. These relaxation dynamics are inconsistent with the disorder-assisted electron-phonon (supercollision) cooling mechanism (see Supplementary Fig. 5-10 and Supplementary Note 6), but they could be attributed to the optical and acoustic phonon cooling mechanisms of hot carriers. 

% IR Experiment
\subsection{Ultrafast degenerate IR pump-probe spectroscopy}
To further support the existence of interlayer energy transfer in MEG, we devise another experiment using ultrafast degenerate IR pump-probe spectroscopy \cite{SunPRL2008, SunPRL2010} in which we selectively photoexcite hot electrons in all layers of MEG except the first HD layer nearest to the SiC substrate. We then observe the cooling of these hot electrons via interlayer Coulomb coupling to the cold electrons in the first HD layer. Our laser system consists again of a Ti:Sapphire oscillator and amplifier followed by an optical parametric amplifier (OPA $9850$, Coherent) and produces ultrafast optical pulses with a center wavelength tuned to $1.8$ $\mu$m. The OPA beam is filtered through a $10$ nm bandpass filter centered at $1.8$ $\mu$m and is split into a pump and a probe beams that are both focused on the MEG sample with an intensity spot size diameter of $\sim 50$ $\mu$m. The transmitted portion of the probe is detected by using a grating spectrometer and an InGaAs photodetector in conjunction with a conventional lock-in amplifier data acquisition technique. The MEG sample is mounted inside a cryostat (ST-100, Janis) and the substrate temperature is held at $10$ K. The experimental error is again estimated not to exceed $\sim 5 \%$. 

\begin{figure}
\begin{center}
\includegraphics[width=4.0in]{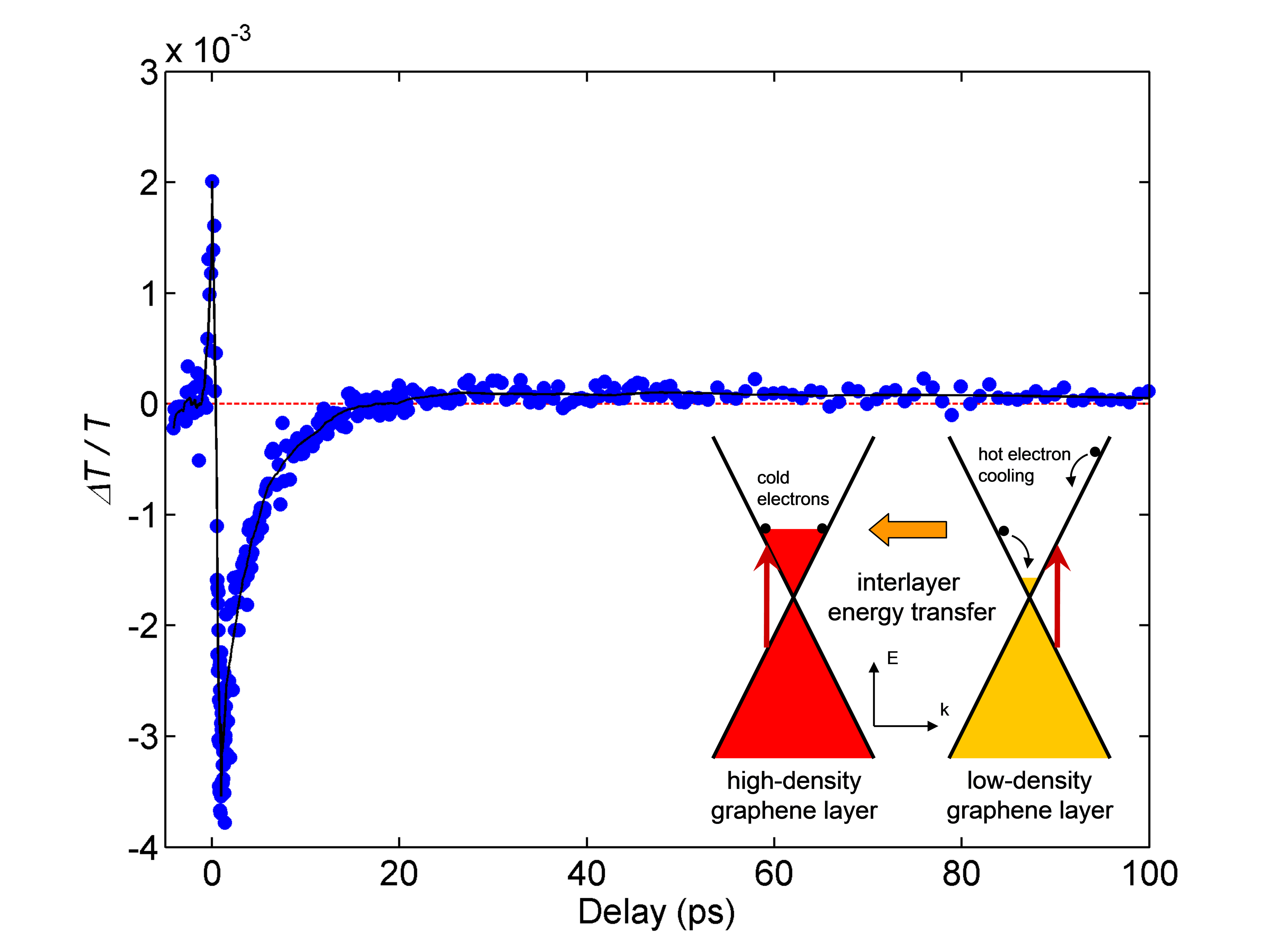}
\caption{{\bf Ultrafast degenerate IR pump-probe spectroscopy on MEG} Normalized probe differential transmission $\Delta T/T$ in ultrafast degenerate $1.8$ $\mu$m IR pump-probe spectroscopy as a function of pump-probe delay recorded at a pump fluence of $80$ $\mu$J cm$^{-2}$ and a substrate temperature of $10$ K for a MEG sample with $\sim 63$ layers. The black solid line is a guide for the eye. The sign changes in the differential transmission at $\sim 1$ and $\sim 20$ ps indicate the presence of interlayer energy transfer from the top layers to the first HD layer in MEG. (Inset: Schematic diagram of interlayer Coulombic energy transfer from a hot LD to a cold HD layer in MEG. The pump selectively injects hot electrons in all layers of MEG \emph{except} the first HD layer, in which interband absorption is Pauli blocked.)}
\label{fig:fig3}
\end{center}
\end{figure}

In this experimental approach, both the pump and the probe photon energies ($\hbar\omega \approx 690$ meV) are chosen to be slightly smaller than twice the Fermi level of the first HD layer ($E_{\rm {F}} \approx 360$ meV), but larger than twice the Fermi levels of all other layers in MEG. \cite{SunPRL2008, SunPRL2010} Thus, the pump selectively injects hot electrons in all layers of MEG \emph{except} the first HD layer, in which interband absorption is Pauli blocked as illustrated schematically in the inset of Figure~\ref{fig:fig3}. The probe differential transmission due to photoexcitation has a \emph{positive} contribution from the hot electrons in all layers of MEG except the first HD layer. The first HD layer gives no contribution, when the carriers in that layer remain unexcited. However, interlayer Coulombic energy transfer can heat this layer, which results in a distinct \emph{negative} contribution to the differential transmission. 

Figure~\ref{fig:fig3} shows the probe differential transmission normalized to the probe transmission without photoexcitation, $\Delta T/T$, as a function of pump-probe delay for the MEG sample with $\sim 63$ layers. We observe that immediately after photoexcitation the differential transmission is positive, arising from the hot electrons that are directly injected in the top layers. Shortly after that, the differential transmission becomes negative and reaches its minimum value within $\sim 1$ ps. This sign change demonstrates the existence of an efficient interlayer energy transfer, in which the cold electrons in the first HD layer act as a heat sink for the hot electrons in the top layers. The rapidly rising electron temperature in the first HD layer has a dominant negative contribution to the differential transmission. As delay time increases, the electrons in the first HD layer cool much faster than the electrons in the top layers due to the sharply increasing rate of acoustic phonon emission with carrier density. This results in a second sign change in the differential transmission at $\sim 20$ ps, when the electron temperature in the first HD layer approaches the equilibrium lattice temperature. By comparing to the THz carrier dynamics in Figure~\ref{fig:fig2}g, we note that it takes slightly longer for the HD layer to cool due to the extra heat from the top LD layers. At that point, the optical and acoustic phonon cooling rate in the first HD layer balances the interlayer energy transfer rate from the top layers. Electronic cooling in the top layers of MEG via the interlayer Coulomb coupling survives on a timescale exceeding a hundred picoseconds (limited by the experimental signal-to-noise ratio), which is consistent with that observed in the THz carrier dynamics in Figure~\ref{fig:fig2}c-d. Now, we turn our attention to a detailed presentation of the theory of hot-carrier equilibration based on interlayer energy transfer via screened Coulomb interactions. 

%% Theory
\subsection{Interlayer Coulombic energy transfer theory}
Non-equilibrium electrons in graphene have been shown to thermalize within a layer on an ultrafast timescale 
on the order of tens of femtoseconds \cite{BreusingPRL2009, BreusingPRB2011, BridaNatComm2013, GilbertsonJPCLett2012, CavalleriNatMat2013}. 
We assume that this property holds also in MEG, and that it leads to 
pseudo-equilibrium electronic states with well defined temperatures $T_{i}$ in layer $i$, but we
allow for the possibility of differences in temperature between layers that survive to longer timescales. 
The multilayer temperature dynamics
are described by a set of coupled non-linear first order differential equations: 
\be
\partial_t T_{i} = \left(\cQ^{\rm {ph}}_{i}(T_{i}) + \sum_{j \ne i}  \cQ^{\rm {el}}_{ij}(T_{i},T_{j})\right)/\cC_i,    \label{eq:Te}
\ee
where $\cC_i = \partial_{T} \cE_i$ is the heat capacity and $\cE_i$ the energy density of electrons in the $i$'th layer.
The rate of change of energy density in layer $i$ is determined by the sum of two processes: energy loss to the lattice via electron-phonon scattering at rate $\cQ^{{\rm ph}}_i$, and energy transfer from other layers $j$ to $i$ via interlayer electron-electron scattering at rate $\cQ^{\rm {el}}_{ij}$.
Both of these mechanisms depend strongly on the carrier density. The four most highly doped layers in the $63$-layer MEG sample have Fermi energies measured to be $360$, $218$, $140$, and $93$ meV, respectively \cite{SunPRL2010}. 
A simple Thomas-Fermi model\cite{Mele} is able to account semi-quantitatively for the monotonic decrease in carrier density with separation from the substrate in multilayer graphene systems. Electronic transport and magneto-optical spectroscopy measurements \cite{deHeerScience2006, PotemskiPRL2011} of top LD layers suggest local carrier density fluctuations in these LD layers that satisfy $n_{\rm {LD}} \lesssim 10^{10}$cm$^{-2}$.

Recent ultrafast optical spectroscopy experiments\cite{SunPRL2008} have found that the hot carriers in the HD layers of MEG quickly relax to the lattice temperature with equilibration times on the order of a few picoseconds, in good agreement with the experiments reported here. Theoretical calculations neglecting interlayer thermal coupling ($\cQ^{\rm {el}}_{ij} \rightarrow 0$) have obtained order of magnitude agreement with these measurements \cite{BistritzerPRL2009}. 
On the other hand, the same approximation applied to the LD layers erroneously predicts thermal equilibration times on the order of several nanoseconds, in sharp disagreement with the experiments reported here. This was the initial impetus for our examination of the interlayer Coulombic energy transfer mechanism. 
According to Ref.~\citenum{BistritzerPRL2009} the acoustic phonon cooling power is proportional to the square of the carrier density, $n^2$.
Allowing for disorder-assisted electron-phonon (supercollision) scattering changes this dependence to $n$ \cite{LevitovPRL2012}. 
In the HD layers, optical and acoustic phonon emission is likely to provide the dominant cooling pathway for hot carriers \cite{SunPRL2008}. 
In the LD layers, however, we suggest that it plays a more subsidiary role by keeping the electron temperature in the HD layers pinned to the lattice temperature ($T_{\rm {HD}}=T_{\rm {L}}$), while they act as a Coulomb-coupled heat sink for the LD layers (see Supplementary Fig. 4 and Supplementary Note 5). 

Although the true electron density profile across the many layers of MEG is expected to decrease smoothly, it is convenient to make a sharp distinction between highly doped (HD) and lightly doped (LD) layers. 
We predict that the asymptotic temperature dynamics of LD layers in MEG are effectively governed by Equation~\ref{eq:Te}
 with $\cQ^{\rm {ph}}_{\rm {LD}} \rightarrow 0$. In our theoretical analysis, we will denote the four most highly doped layers near the substrate, as HD layers. Quantitatively, this cutoff is suggested from the disorder-free theory of acoustic phonon cooling \cite{BistritzerPRL2009}, which 
in combination with Equation 4 in Supplementary Note 2, 
implies that the hot-electron distribution in layers $i>4$ transfers energy via the interlayer Coulomb interaction to 
the $j$ HD layers ($j<i$) faster than it loses energy to the lattice via acoustic phonon emission. 
More precisely, we find that  $\cQ^{\rm {ph}}_i/(\sum_{j<i}\cQ^{\rm {el}}_{ij}) \lesssim 1$, for temperatures $T_i \gtrsim 50$ K. 
In all calculations described below, we use the values $E_{{\rm F},1-4}=360$, $218$, $140$, and $93$ meV for the Fermi levels of the first four HD layers. 
These have been measured explicitly for the $63$-layer MEG sample and are expected to be good estimates for the $35$-layer MEG sample. 

The rate of Coulombic energy transfer between two layers is given by a Fermi golden-rule expression: 
\begin{eqnarray}\label{eq:eratesimp}
 \cQ^{\rm {el}}_{ij} &=& - \frac{ 2 \pi }{\hbar} \sum_{ \bm{k_i} ,  \bm{k_i^{\prime}}} \sum_{ \bm{k_j} ,  \bm{k_j^{\prime}}}  \left( \ve_{\bm{k_j^{\prime}}} - \ve_{\bm{k_j}} \right) |\cW_{\rm {int}}|^2 \nonumber \\
 &\times &  \delta_{\bm{k}_{i'}+\bm{k}_{j'},\bm{k}_{i}+\bm{k}_{j}} \; \delta( \ve_{\bm{k}_{j'}} + \ve_{\bm{k}_{i'}} - \ve_{\bm{k}_j} - \ve_{\bm{k}_i} ) 
 \nonumber \\
 &\times& \left( f_{\bm{k_j}}\left( 1 - f_{\bm{k_j^{\prime}}} \right) f_{\bm{k_i}}\left( 1 - f_{\bm{k_i^{\prime}}} \right)  -   f_{\bm{k_j}^{\prime}}\left( 1 - f_{\bm{k_j}} \right) f_{\bm{k_i}^{\prime}}\left( 1 - f_{\bm{k_i}} \right) \right),  
\end{eqnarray}
\noindent
where $(i,j)$ are layer indices, and $\bm{k}$ is a collective index
which implies, in addition to wavevector, the spin, the valley, and the band index labels 
required to specify single electron states in graphene's low energy Dirac model\cite{Semenoff}.
As mentioned above, \emph{intra}-layer electron thermalization \cite{BreusingPRL2009, BreusingPRB2011, BridaNatComm2013, GilbertsonJPCLett2012, CavalleriNatMat2013} is much faster than \emph{inter}-layer energy transfer, and this allows us to describe electronic state occupations with a quasi-equilibrium Fermi distribution $ f_{\bm{k}}= f( \ve_{\bm{k}},\mu,T)$. In the random phase approximation (RPA): 
\begin{equation}\label{eq:amplitude}
\begin{split}
\cW_{\rm {int}} = & \left(\frac{1 + \alpha_i \beta_i e^{i(\theta_{\bm{k_i}-\bm{q}}-\theta_{\bm{k_i}})}}{2}\right) \left(\frac{ (1 + \alpha_j \beta_j e^{i(\theta_{\bm{k_j}+\bm{q}}-\theta_{\bm{k_j}})}}{2}\right) \\
& \times v^{\rm {sc}}_{ij}(q,\omega) \;  \delta_{\sigma_i,\sigma_i^{\prime}} \delta_{\sigma_j,\sigma_j^{\prime}} \delta_{\tau_i,\tau_i^{\prime}} \delta_{\tau_j,\tau_j^{\prime}},
\end{split}
\end{equation}
\noindent
where $q=|\bm{k}_{j'}-\bm{k}_{j}|$, $\epsilon= \ve_{\bm{k}_{j'}}  - \ve_{\bm{k}_j}$, and 
$v^{\rm {sc}}_{ij}(q,\epsilon)$ is the screened electron-electron interaction between
layers $i$ and $j$ (see below).  When screening is neglected,
$v^{\rm {sc}}_{ij}(q,\epsilon) \rightarrow v_{ij}(q) = 2\pi e^2 \exp(-q d_{ij})/ \kappa q$ is the 2D
Fourier transform of the bare Coulomb interaction
between two electrons with interlayer separation $d_{ij}$ 
and $\kappa = 5.5$ to account for the presence of MEG at the surface of the SiC substrate.  
The factors in parenthesis are the well-known form factors that account for the sub-lattice spinor  
dependence of graphene $\pi$-band plane-wave matrix elements. 
The indices $\alpha$ and $\beta$ are equal to $1$ and $-1$ for conduction and valence band states
respectively.  The Kronecker delta's in Equation~\ref{eq:amplitude} explicitly exhibit the property that 
continuum model interactions are independent of 
spin ($\sigma$) and valley ($\tau$).   Using Equation~\ref{eq:amplitude} and comparing with 
the expression for  graphene's non-interacting density-density response function, $ \chi(q,\omega,T) $,
we finally obtain the following compact expression which is suitable for numerical evaluation: 
\begin{equation}\label{eq:erate}
\begin{split}
\cQ^{\rm {el}}_{ij} = & \frac{\hbar}{\pi}\int^{\infty}_{- \infty}  \, d\omega \; \omega \sum_{\vec{q}}| v^{\rm {sc}}_{ij}|^2 \\
& \times [ n_{\rm B}(\hbar \omega/T_i)- n_{\rm B}(\hbar \omega/T_j)] \\
& \times {\rm Im}[\chi_i (q,\omega,T_i)] \, {\rm Im}[\chi_j (q,\omega,T_j)],
\end{split}
\end{equation}
\noindent
where $n_{\rm B}(x)=1/(\exp(x)-1)$ and $k_{\rm B}=1$ throughout.  We use this expression below to calculate interlayer energy transfer rates. 
A central quantity in the theoretical formulation of the many-body effects of Dirac fermions is the dynamical polarizability tensor $\chi_i (q,\omega,T_i)$ for the $i$'th layer at temperature $T_i$. This is defined through the one-body non-interacting Green's functions \cite{Gonzalez}. The density-density response function of the doped 2D Dirac electron model was first considered by Shung \cite{Shung} at zero-temperature as a step toward a theory of collective excitations in graphite. The Dirac electron expression $\chi_i (q,\omega,T_i)$ at finite temperature has been recently considered~\cite{Asgari,Faridi,Tomadin}. 
Before proceeding with the rate calculations, however, we must first discuss the approximation we use for the screened interlayer Coulomb interaction. 

In the RPA, the bare interlayer Coulomb interaction is screened\cite{Vignale} by the potential
produced by self-consistently readjusted charge density.
For a general multilayer system this implies that\cite{Profumo}: 
\begin{equation} \label{eq:Vsc}
\bm{v}^{\rm {sc}}_{ij} = \big(\bm{v}^{-1} -  \delta_{i,j} \, \chi_i (q,\omega,T_i) \big)^{-1}_{ij},
\end{equation}
where $\bm{v}$ and $\bm{v}^{\rm {sc}}$ are matrices which describe bare and 
screened potentials in one layer ($i$) to an external charge in another layer ($j$).
Screening complicates energy cooling dynamics, because it causes the
energy transfer rate between a particular pair of layers to depend, through $\chi(q,\omega,T)$, on the 
temperatures in all layers.  However, the low carrier densities in the layers far from the substrate  \cite{deHeerScience2006, deHeerPNAS2011}
motivate a simplifying approximation in which their contributions to screening
are neglected.  By comparing this approximation with the full expression, we find that this simplification
is justified in the regions of phase space ($q,\omega$) important for low temperature energy transfer. 

The relative ability of intraband excitations in LD and HD layers to screen the interaction can be
established by examining the ratio of their density response functions in the
static limit: $\chi_{\rm {HD}}(q)/\chi_{\rm {LD}}(q) = \sqrt{n_{\rm {HD}}/n_{\rm {LD}}}$. In our MEG samples this ratio varies within $\sim 8-31$ in the important regions, and suggests that the leading order charge polarization responsible for screening the MEG interlayer Coulomb interaction can be approximated without
including the contribution from the LD layers. 
We note that the Bose factors in Equation~\ref{eq:erate} limit transfer energies to $\hbar \omega \lesssim k_{\rm B}T$.  
We also note that the factor $\exp(-q d_{ij})$ in the interlayer Coulomb interaction limits the important 
wavevector transfers to
$q < 1/d_{ij}$.  The conditions (setting $\hbar$ and $k_{\rm B} \to 1$), 
\be
q \lesssim 1/d_{ij}, \qquad  \omega \lesssim T,
\label{eq:restrictions}
\ee
\noindent
apply equally well to Coulomb-mediated interlayer energy and momentum transfer 
in any type of multilayer 2D electron system. 

Setting $\chi_j(q,\omega,T_j)$ to zero for all but the four HD layers nearest the substrate, and letting the separation distance between the four HD layers also go to zero (relative to the generally much larger spacing between HD and LD layers in MEG), we obtain the following approximate expression: 
\begin{equation} 
v^{\rm {sc}}(q,\omega)_{ij} \to v_{ij}(q)/\epsilon^{\rm {MEG}}(q,\omega),
\end{equation}
where 
\begin{equation}
\epsilon^{\rm {MEG}}(q,\omega)=1-\frac{2\pi e^2}{\kappa q}\sum_{j \in {\rm HD}} \chi_{j}(q,\omega,T_j).
\label{eq:dielectric}
\end{equation}
\noindent
In the calculations reported on below $j$ was summed over the four HD layers. 

Since cooling powers between LD layers are typically $1$ to $2$ orders of magnitude larger than
between LD and HD layers \cite{Soljacic} ($\cQ^{\rm {el}}_{\rm {LD, LD^{\prime}}} \gg \cQ^{\rm {el}}_{\rm {LD, HD}}$), 
energy transfer between pairs of LD layers
cannot be ignored in the calculations.   Poles in the screened interaction described by Equation~\ref{eq:Vsc} at plasmon frequencies greatly enhance the interlayer quasi-particle scattering rate. This effect does not contribute to energy transfer between HD and LD layers, because the plasmon modes then
lie at higher frequencies than can be excited in low temperature HD layers; the plasmon poles reside above the \emph{intra}-band particle-hole continuum of HD layers, {\em i.e.} $\omega_{\rm {pl}}(q) > v_{\rm F} q $. 
However, at temperatures comparable to the Fermi temperature, \emph{inter}-band particle-hole excitations are no longer Pauli blocked at any frequency, and these excitations can take advantage of the plasmon poles in the screened interaction \cite{Hill1997, Flensberg1995}. It is the 
small Fermi temperatures ($T_{\rm {F}} \propto \sqrt{n}$) of the LD layers 
that dramatically increases the Coulomb-coupling amongst the group of LD layers.

\begin{figure}
\begin{center}
\includegraphics[width=4.0in]{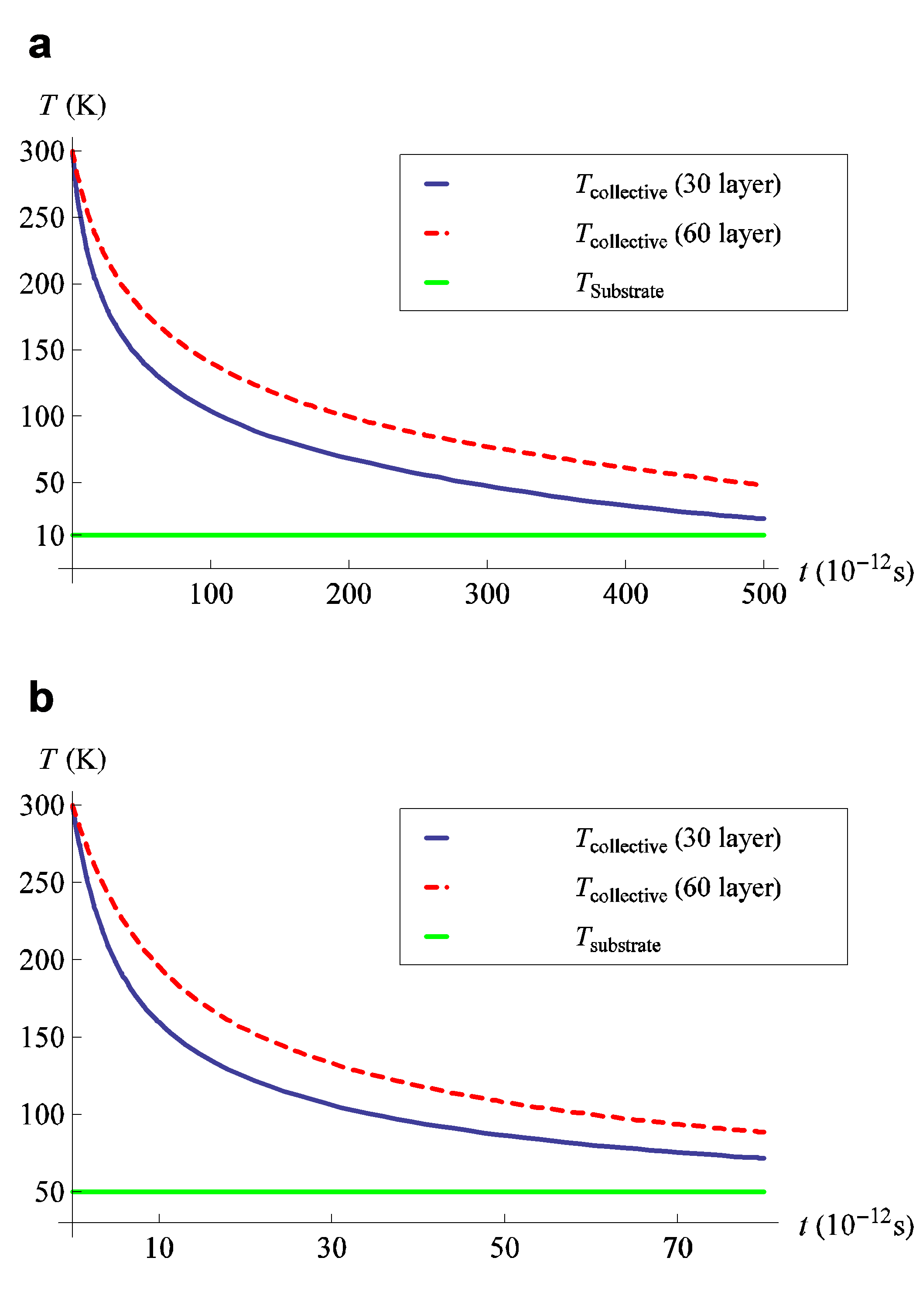}
\caption{{\bf Interlayer Coulombic energy transfer in MEG} Collective cooling dynamics of the LD layers in MEG with 30 and 60 layers, calculated using Equation~\ref{eq:collcool}. Subfigure (a) is for lattice temperature $T_{\rm L}=10$ K and Subfigure (b) is for lattice temperature $T_{\rm L}=50$ K. Both figures illustrate that the MEG hot-carrier relaxation characteristics observed in Figure~\ref{fig:fig2} are naturally described by the interlayer Coulombic energy transfer mechanism. Samples with more LD layers cool more slowly because Coulomb coupling to the HD layers close to the substrate decreases with increasing layer separation.}
\label{fig:fig4}
\end{center}
\end{figure}

As LD layers closer to the substrate cool, they absorb energy from and cool more 
distant LD layers.    
This effect results in a \emph{collective} cooling state in which the electron temperatures of all LD layers relax to the equilibrium lattice temperature 
nearly uniformly, and motivates an approximation with a single collective LD layer temperature, $T_{\rm {c}}(t)$. In employing this approximation our goal is to establish that Coulomb scattering is a relevant energy transfer process up to quite high temperatures.  A more detailed calculation in which the temperature of each LD layer is allowed to vary independently would be warranted if the charge density profile in the multilayer system was accurately known.  Such a calculation might in any event not achieve greater accuracy since the calculations of electron-electron scattering amplitudes can only be performed approximately.  The collective cooling state model we employ has the advantage that it requires fewer computationally burdensome finite-temperature dynamic polarizability calculations. The collective cooling dynamics model of the collective temperature is given by: 
\be
\partial_t T_{\rm {c}} = \left(\sum_{i \in {\rm LD}} \sum_{j \in {\rm HD}} \cQ^{\rm {el}}_{i j}(T_{\rm {HD}}=T_{\rm {L}},T_{\rm {LD}}\rightarrow T_{\rm {c}},d_{ij})\right)/N \cC_{\rm {LD}},   \label{eq:collcool}
\ee
\noindent
where $N$ is the total number of LD layers. We have summed over the energy transfer rate between all LD-HD pairs of layers, appealing to 
strong electron-phonon coupling to keep the HD layers temperature at $T_{\rm L}$ and to strong $\cQ^{\rm {el}}_{\rm {LD, LD^{\prime}}}$ to 
keep the LD layers at a common temperature $T_{\rm c}(t)$.  Figure~\ref{fig:fig4}a-b compares the calculated collective thermal relaxation 
dynamics $T_{\rm c}(t)$ in 30-layer and 60-layer MEG samples for lattice temperature of 10 K and 50 K. 
For both lattice temperatures, thicker MEG samples cool more slowly. 
This trend can be understood by noting that each additional LD layer contributes an equal amount to the combined LD layers
heat capacity, whereas the energy transfer rate to the HD layers falls off upon moving further away from the substrate (see Supplementary Fig. 3 and Supplementary Note 4).

\begin{figure}
\begin{center}
\includegraphics[width=4.0in]{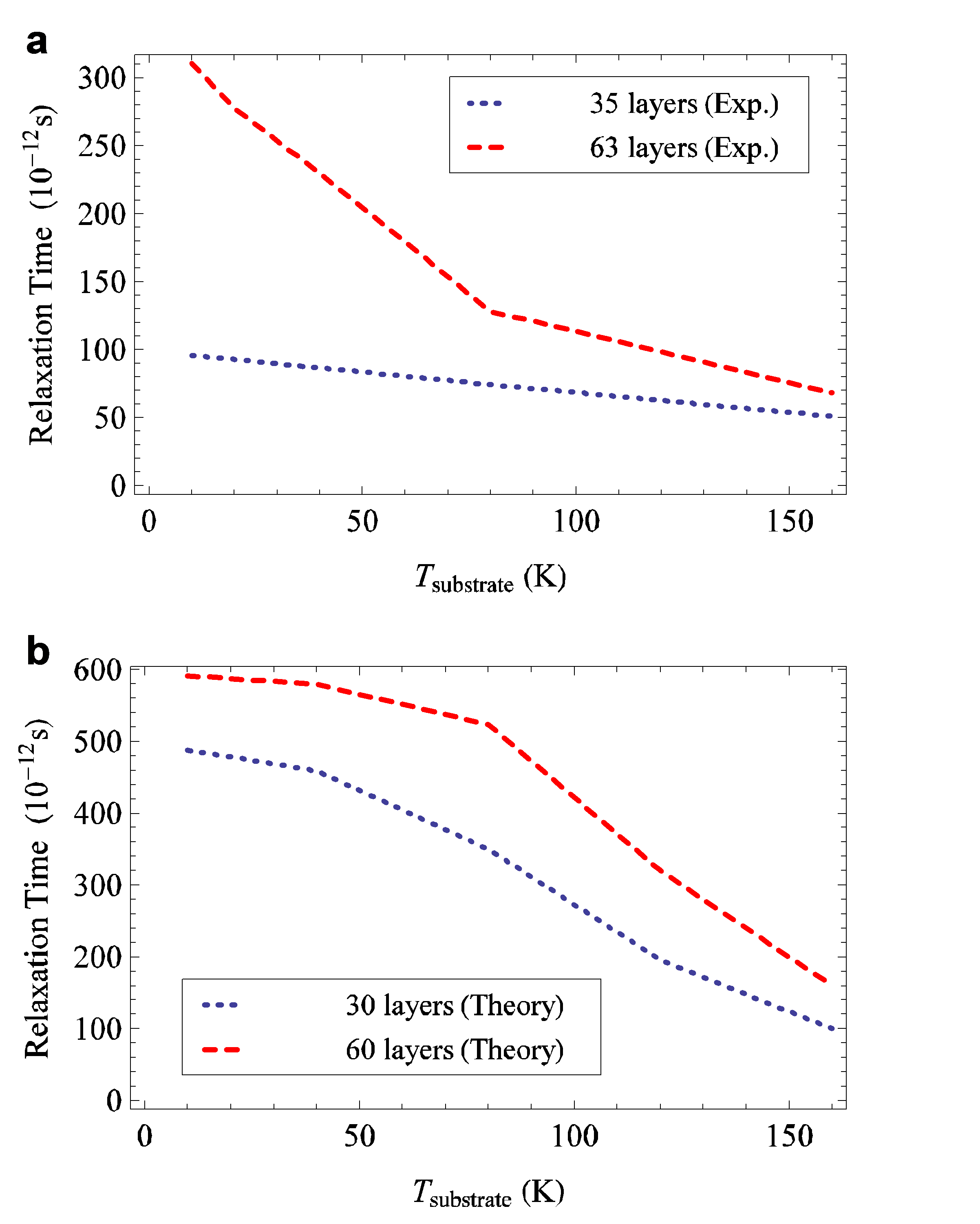}
\caption{{\bf Comparison between experimental and theoretical relaxation times in MEG} Subfigure (a) shows experimental hot-carrier relaxation times versus substrate temperature for MEG samples with $\sim 63$ and $\sim 35$ layers. The values correspond to the long relaxation times from Figure~\ref{fig:fig2}e-f. Subfigure (b) shows theoretical hot-carrier relaxation times versus substrate temperature calculated using the linearized interlayer Coulombic energy transfer rate Equation~\ref{eq:eratediff}. The interlayer Coulombic energy transfer mechanism explains relaxation time trends versus substrate temperature and multilayer system thickness. The factor of two discrepancy in magnitude is typical for RPA theories of electron-electron scattering amplitudes. The level of agreement found here is similar to that found in comparisons of RPA calculations and measurements of the Coulomb-coupled interlayer momentum transfer rate (i.e. Coulomb drag resistivity) between neighboring GaAs/AlGaAs quantum wells \cite{Hill1997}.}
\label{fig:fig5}
\end{center}
\end{figure}

If we linearize the collective temperature of the LD layers about the equilibrium lattice temperature, $\delta T \equiv T_{\rm c}(t)-T_{\rm L}$, we find that the interlayer energy transfer rate can be written as: 
\begin{equation}\label{eq:eratediff}
\begin{split}
\cQ^{\rm {el}}_{ij} = & \frac{\hbar}{\pi}\int^{\infty}_{- \infty}  \omega d\omega \sum_{\vec{q}}| v^{\rm {sc}}_{ij}|^2 \\
& \times \frac{\hbar \omega}{4 T_{\rm L}^2}\frac{\delta T}{\sinh^2(\hbar \omega / 2 T_{\rm L})} \\
& \times {\rm Im}[\chi_i (q,\omega,T_{\rm L})] {\rm Im}[\chi_j (q,\omega,T_{\rm L})].
\end{split}
\end{equation}
\noindent
This equation can be used to define collective electronic cooling times which are easier to compare to the experimental carrier relaxation times extracted from the phenomenological fits. 
Figure~\ref{fig:fig5} compares the experimental and the theoretical relaxation times over a lattice temperature range of $T_{\rm L} = 10-160$ K. From this figure we see that, within our approximations, theory reproduces the experimental trends versus both the lattice temperature and the number of epitaxial graphene layers; more specifically, both relaxation times increase with decreasing lattice temperature and increasing the number of layers. We note that since our theoretical relaxation times are obtained by linearizing the interlayer energy transfer rate equations, they underestimate interlayer coupling except when $T_{\rm {LD}}-T_{\rm {HD}} \ll T_{\rm {HD}}$. 
Thus, we expect (and observe) the theoretical relaxation times to be longer than those obtained by fitting the experimental data over a broader temperature range. 

We can identify the microscopic origins of the relaxation times' dependence on lattice temperature using a formula derived in Supplementary Note 2, where 
we give an analytic formula for the interlayer Coulombic energy transfer rate between a single LD layer and a single HD layer in MEG. 
Here, we show only the result of summing over the rates between all HD-LD pairs of layers. 
Applying this result to the multilayer case, we 
obtain the following net energy loss rate (per area) for $N$ LD layers in MEG: 
\begin{equation}\label{eq:DegenErate}
\begin{split}
\frac{1}{L^2} \cdot \cQ^{\rm {el}}_{\rm {HD, LD}} =  & \gamma (T_{\rm {LD}}-T_{\rm {HD}}) \left( \frac{N \nu_{\rm {LD}}}{(\sum_{j}\nu_{\rm {HD}})} \right) \\
& \times  T_{\rm {HD}}^3 \ln \left( \frac{T_{\rm {F, LD}}}{T_{\rm {HD}}} \right),
\end{split}
\end{equation}
\noindent
where $\gamma = (8 \pi^2 / 30)(k_{\rm B}^4/\hbar^3 v_{\rm F}^2)$ and 
the density of states in the HD and LD layers are denoted by $\nu_{\rm {HD}}$ and $\nu_{\rm {LD}}$. 
The sum over index $j$ runs over all HD layers in the MEG sample. We find that the interlayer Coulombic energy transfer rate exhibits a temperature dependence of $T^3 \ln (T)$. \cite{SongEnergy} 
This result is related to the familiar result\cite{Vignale} for electron-electron scattering rates $\tau^{-1}$ in Fermi liquids which are proportional to $ T^2 \ln(T)$.  These power laws appear, because the Fermi distribution limits the initial-final state pair energies which can partake in scattering to an energy window of 
width $\hbar \omega \sim k_{\rm B} T$ about the Fermi energy, which qualitatively explains the additional factor of $T$; the additional factor of $T$ appears because scattering events are
weighted by the transferred energy in the cooling power case. 
On the other hand, the heat capacity of nearly neutral graphene increases linearly with $T$ (or as $T^2$ if $T \ll T_{\rm F}$). 
The final result is that the electronic cooling time defined above becomes shorter with increasing lattice temperature, consistent with the experimental trends as shown in Figure~\ref{fig:fig5}.
Precise agreement is not expected outside of the degenerate temperature regime ($T_{\rm L} \ll T_{\rm {F,LD}} \sim 135$ K), within which Equation~\ref{eq:DegenErate} becomes exact.

%% Conclusion
%\section{Conclusion}
\section{Discussion}
In conclusion, we have developed a theory of hot-carrier equilibration based on interlayer energy transfer via screened Coulomb interactions between electrons in the many top low-density (LD) layers and in the few high-density (HD) layers close to the underlying SiC substrate in multilayer epitaxial graphene (MEG). The theory is complicated by the essential role of dynamic screening of the electron-electron interactions in all MEG layers through the temperature-dependent charge carrier response. To obtain a transparent theory, we have made two well-justified simplifications. First, we note that screening in the relevant temperature, wave vector, and frequency regime is dominated by the first few HD layers close to the substrate, allowing us to neglect the screening by the top LD layers. Second, we note that interlayer energy transfer among the LD layers is much stronger than between LD and HD layers, and we therefore can describe all LD layers by a common electron temperature. We have compared the calculated cooling dynamics with the relaxation dynamics measured via ultrafast time-resolved THz spectroscopy. The observed experimental dynamics exhibit the expected timescales, dependence on lattice temperature, and dependence on number of epitaxial graphene layers predicted by the theory, providing strong support for the proposed mechanism, within the approximations necessary in the development of the theory. The theoretical approach developed here may be expected to be applicable to many other types of layered 2D electron systems. These may include semiconductor heterostructures as well as the wide variety of novel 2D materials under active development including transition metal dichalcogenides (e.g. MoS$_{2}$, MoSe$_{2}$, WS$_{2}$, WSe$_{2}$, etc.) \cite{StranoNatNano2012}, and other van der Waals heterostructures \cite{GeimNature2013, BritnellScience2013}. 

%%%%%%%%%%%%%%%%%%%%%%%%%%%%%%%%%%%%%%%%%%%%%%%%%%%%%%%%%%%%%%%%%%%%%
%% Figures
%%%%%%%%%%%%%%%%%%%%%%%%%%%%%%%%%%%%%%%%%%%%%%%%%%%%%%%%%%%%%%%%%%%%%

%% Figure 1

%% Figure 2

%% Figure 3

%% Figure 4

%% Figure 5

%%%%%%%%%%%%%%%%%%%%%%%%%%%%%%%%%%%%%%%%%%%%%%%%%%%%%%%%%%%%%%%%%%%%%
%% Methods
%%%%%%%%%%%%%%%%%%%%%%%%%%%%%%%%%%%%%%%%%%%%%%%%%%%%%%%%%%%%%%%%%%%%%

%%%%%%%%%%%%%%%%%%%%%%%%%%%%%%%%%%%%%%%%%%%%%%%%%%%%%%%%%%%%%%%%%%%%%
%% References
%%%%%%%%%%%%%%%%%%%%%%%%%%%%%%%%%%%%%%%%%%%%%%%%%%%%%%%%%%%%%%%%%%%%%
\newpage
%\section{References}

%%%%%%%%%%%%%%%%%%%%%%%%%%%%%%%%%%%%%%%%%%%%%%%%%%%%%%%%%%%%%%%%%%%%%
%% Acknowledgements
%%%%%%%%%%%%%%%%%%%%%%%%%%%%%%%%%%%%%%%%%%%%%%%%%%%%%%%%%%%%%%%%%%%%%
\section{Acknowledgements}
M. T. M., C. J. D. and D. S. thank Steve Katnik for technical assistance with the ultrafast laser system. 
The work at the University of Michigan and the Georgia Institute of Technology was supported in part by the National Science Foundation (NSF) Materials Research Science and Engineering Center (MRSEC) under grant DMR-0820382 and DMR-1120923. 
The work at the University of Texas was supported in part by the Welch Foundation under grant TBF1473 and the Department of Energy (DOE) Division of Materials Sciences and Engineering under grant DE-FG03-02ER45958. 
M. P. acknowledges partial financial support from the European Commission (EC) under the Graphene Flagship program (contract No. CNECT-ICT-604391) and MIUR through the program ``Progetti Premiali 2012'' (project ``ABNANOTECH''). 
C. B. and W. A. d. H. acknowledge partial financial support from the AFSOR, and C. B. acknowledges partial financial support from the EC under the Graphene Flagship program (contract No. CNECT-ICT-604391). 

%%%%%%%%%%%%%%%%%%%%%%%%%%%%%%%%%%%%%%%%%%%%%%%%%%%%%%%%%%%%%%%%%%%%%
%% Author contributions
%%%%%%%%%%%%%%%%%%%%%%%%%%%%%%%%%%%%%%%%%%%%%%%%%%%%%%%%%%%%%%%%%%%%%
\section{Author contributions}
M. T. M. and J. R. T. contributed equally. 
M. T. M. and C. J. D. performed and analyzed the ultrafast time-resolved THz spectroscopy experiments. 
D. S. performed the ultrafast degenerate IR pump-probe spectroscopy experiments. 
J. R. T., R. A., M. P. and A. H. M. developed the theory of interlayer energy transfer via screened Coulomb interactions. 
M. T. M. organized the discussion on disorder-assisted electron-phonon (supercollision) cooling. 
C. B. and W. A. d. H. provided the multilayer epitaxial graphene (MEG) samples. 
M. T. M., J. R. T., A. H. M. and T. B. N. wrote the paper. 
All authors discussed the results. 

%%%%%%%%%%%%%%%%%%%%%%%%%%%%%%%%%%%%%%%%%%%%%%%%%%%%%%%%%%%%%%%%%%%%%
%% Additional information
%%%%%%%%%%%%%%%%%%%%%%%%%%%%%%%%%%%%%%%%%%%%%%%%%%%%%%%%%%%%%%%%%%%%%
\section{Additional information}
Supplementary information is available in the online version of the paper. Reprints and permissions information is available online at www.nature.com/reprints. Correspondence should be addressed to T. B. N. (tnorris@umich.edu). 

%%%%%%%%%%%%%%%%%%%%%%%%%%%%%%%%%%%%%%%%%%%%%%%%%%%%%%%%%%%%%%%%%%%%%
%% Competing financial interests
%%%%%%%%%%%%%%%%%%%%%%%%%%%%%%%%%%%%%%%%%%%%%%%%%%%%%%%%%%%%%%%%%%%%%
\section{Competing financial interests}
The authors declare no competing financial interests.

%%%%%%%%%%%%%%%%%%%%%%%%%%%%%%%%%%%%%%%%%%%%%%%%%%%%%%%%%%%%%%%%%%%%%
%% Supplementary Information
%%%%%%%%%%%%%%%%%%%%%%%%%%%%%%%%%%%%%%%%%%%%%%%%%%%%%%%%%%%%%%%%%%%%%
\newpage
\section{Supplementary Information}

\setcounter{equation}{0}
\setcounter{figure}{0}
\renewcommand{\figurename}{Supplementary Figure}

%%%%%%%%%%%%%%%%%%%%%%%%%%%%%%%%%%%%%%%%%%%%%%%%%%%%%%%%%%%%%%%%%%%%%
%% Supplementary Notes
%%%%%%%%%%%%%%%%%%%%%%%%%%%%%%%%%%%%%%%%%%%%%%%%%%%%%%%%%%%%%%%%%%%%%
%\section{Supplementary Notes}

%% Supplementary Note 1
%\newpage
\subsection{Supplementary Note 1 \\ Normalized differential THz transmission spectra $\Delta t(\omega)/t(\omega)$}
\begin{figure}
\begin{center}
\includegraphics[width=4.0in]{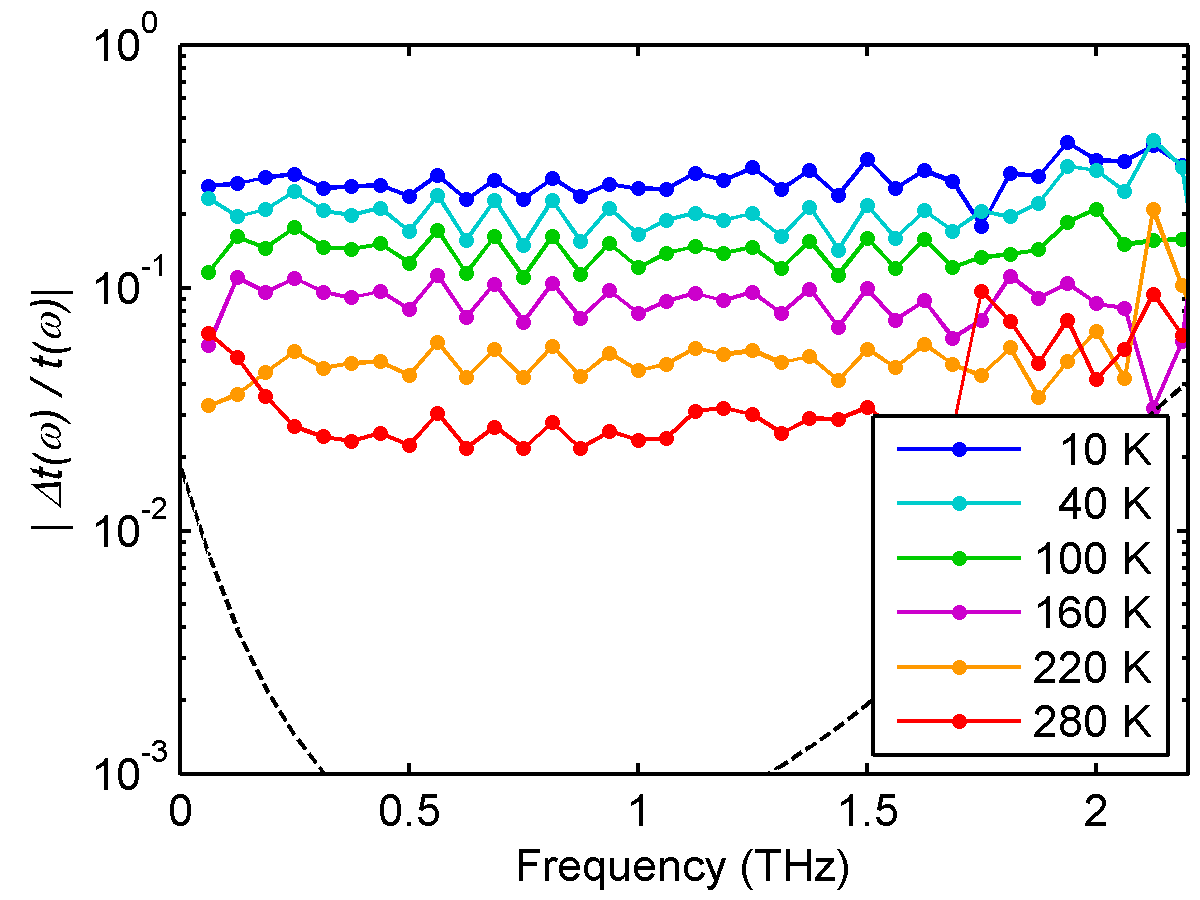}
\caption{{\bf Ultrafast time-resolved THz spectroscopy on MEG.} 
Normalized differential THz transmission spectra $\Delta t(\omega)/t(\omega)$ recorded at a pump fluence of $0.87$ $\mu$J cm$^{-2}$ and a pump-probe delay of $1$ ps for a few different substrate temperatures for a MEG sample with $\sim 63$ layers. The black dashed line indicates the experimental noise level.}
\label{fig:figS1}
\end{center}
\end{figure}

\begin{figure}
\begin{center}
\includegraphics[width=4.0in]{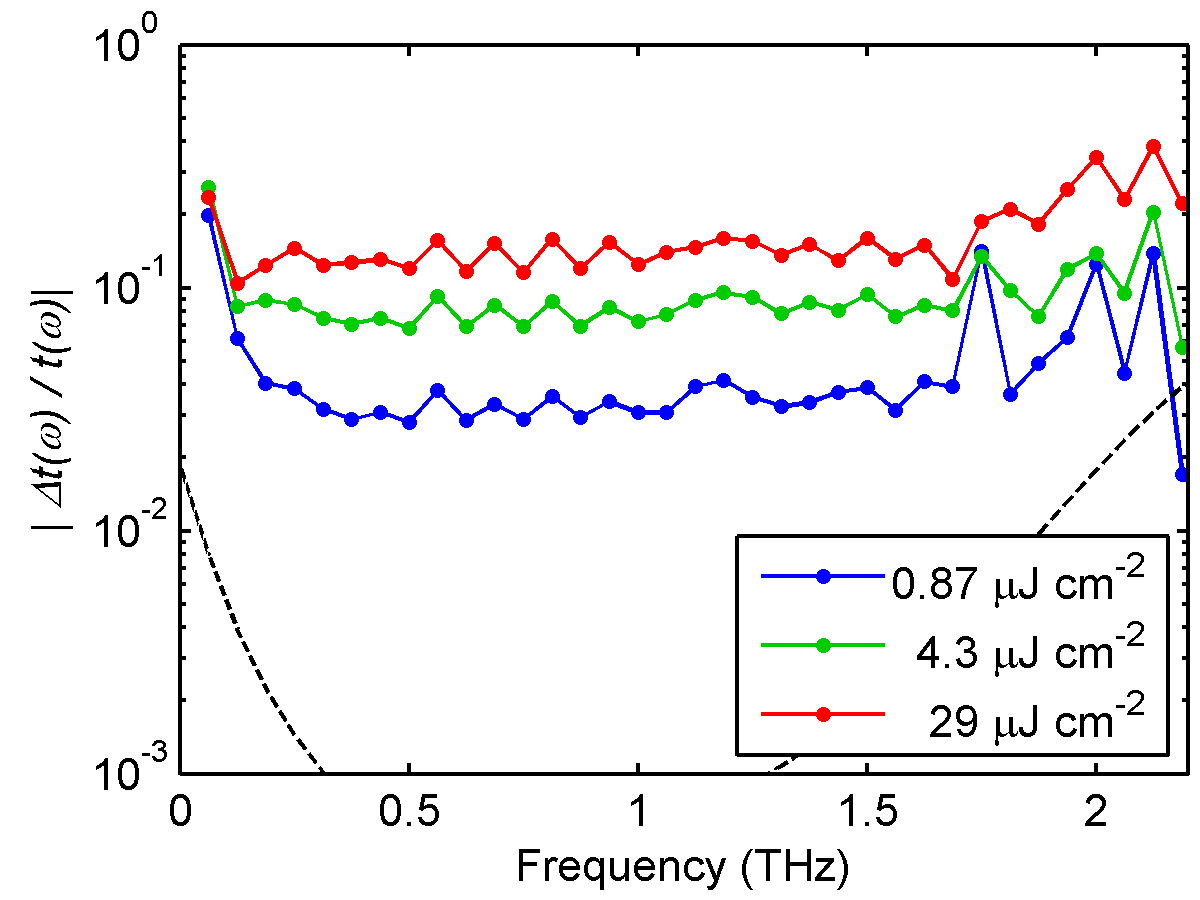}
\caption{{\bf Ultrafast time-resolved THz spectroscopy on MEG.} 
Normalized differential THz transmission spectra $\Delta t(\omega)/t(\omega)$ recorded at a substrate temperature of $280$ K and a pump-probe delay of $1$ ps for a few different pump fluences for a MEG sample with $\sim 63$ layers. The black dashed line indicates the experimental noise level.}
\label{fig:figS2}
\end{center}
\end{figure}

Here, we present additional experimental data to further demonstrate that the normalized differential THz transmission spectra are remarkably dispersionless in the detectable frequency range under all experimental conditions. Supplementary Figure~\ref{fig:figS1} and Supplementary Figure~\ref{fig:figS2} show the differential THz transmission spectra normalized to the THz transmission without photoexcitation, $\Delta t(\omega)/t(\omega)$, for variable substrate temperature and variable pump fluence, respectively, for a MEG sample with $\sim 63$ layers. The frequency-independent dynamic THz response of the MEG samples justifies recording the normalized differential THz transmission only at the peak of the THz probe pulse, $\Delta t/t$, as a function of pump-probe delay to map out the relaxation dynamics of the photoexcited carriers. The slight fluctuations in a few of the data scans for frequencies $\gtrsim 1.7$ THz are due to water vapor absorption arising from very slight fluctuations in the humidity level between the sample and the reference scans. We note that spectra obtained from Fourier-transformed time-domain measurements are much more susceptible to noise than direct spectrally-resolved measurements, because (long-term) fluctuations in the time domain are transferred into uncertainties in the frequency domain during the Fourier transformation process. On the other hand, direct time-resolved THz spectroscopy measurements of the relaxation dynamics are less prone to noise from fluctuations, and a very high signal-to-noise ratio (SNR) can be achieved through sufficient integration. 

%% Supplementary Note 2
\newpage
\subsection{Supplementary Note 2 \\ Interlayer energy transfer linearized in $\Delta T$}
Here, we derive an approximation for the interlayer Coulombic energy transfer rate between a lightly doped (LD) layer and a highly doped (HD) layer in MEG systems. We neglect electron tunneling between layers. Energy transfer between layers can nevertheless occur when an electron in one layer scatters off an electron in a remote layer. To first order in the temperature difference $\delta T = T_{\rm{LD}}-T_{\rm{HD}}$, we find that the energy transfer rate between thermal electron distributions in two Coulomb-coupled layers is given by: 
\begin{equation}\label{eq:eratediffappend}
\begin{split}
\cQ^{\rm{el}}_{ij} = & \frac{\hbar}{\pi}\int^{\infty}_{- \infty}  \omega d\omega \sum_{\vec{q}}| v^{\rm{sc}}_{ij}|^2 \\
& \times \frac{\hbar \omega}{4 T_{\rm L}^2}\frac{\delta T}{\sinh^2(\hbar \omega / 2 T_{\rm{HD}})} \\
& \times {\rm Im}[\chi_i (q,\omega,T_{\rm{HD}})] {\rm Im}[\chi_j (q,\omega,T_{\rm{HD}})].
\end{split}
\end{equation}
\noindent
Equation~\ref{eq:eratediffappend} can be derived by summing over all interlayer electron collision processes and combining a Fermi golden-rule expression for the transition rates with a random phase approximation expression for the electron-electron scattering amplitudes. We focus on the low temperature limit (i.e. $T \ll T_{\rm{F,LD}}, T_{\rm{F,HD}}$), where several physical approximations can be made. In this case, it is clear that the $\omega \rightarrow 0$ limit dominates the integrand of Equation~\ref{eq:eratediffappend}. We note that in this low frequency limit, ${\rm Im}[\chi_i(q,\omega \rightarrow 0)] = -\nu_i \omega / (v_{\rm F} q)$, where $\nu_i = 2 E_{{\rm F},i}/(\pi \hbar^2 v_{\rm F}^2)$ is the density of states at the Fermi energy. This linear dependence on frequency is also found in parabolic band 2DEG's, and describes how the particle-hole excitation spectrum vanishes with decreasing excitation energy $\hbar \omega $  \cite{VignaleSI}. Additionally, in the degenerate regime interlayer particle scattering allows a maximum change in electron wavevector of $\Delta q_{\rm{max}} = 2 k_{\rm F}$, a fact reflected in: 
\begin{equation}
 {\rm Im}[\chi_i(q ,\omega \rightarrow 0)]=0,   \qquad  q > 2 k_{\rm F}.
\label{eq:imchi}
\end{equation}
\noindent
The interlayer Coulomb interaction, proportional to $e^{-q d}$, naturally places the limit  $q \lesssim 1/d$, where $d$ is the interlayer separation. However, $k_{\rm{F,HD}} \ll 1/d$ and we can approximately neglect interlayer separation (i.e. $d \rightarrow 0 $) in Equation~\ref{eq:eratediffappend}. Then, the MEG dielectric function reduces to a Thomas-Fermi-like form: 
\begin{equation}
\epsilon^{\rm{MEG}}_{\rm{TF}}(q)=1+\frac{q^{\rm{MEG}}_{\rm{TF}}}{q}, \qquad q^{\rm{MEG}}_{\rm{TF}}= \frac{2 \pi e^2}{\kappa}\sum_{j} \nu_j,
\label{eq:appenddielectric}
\end{equation}
\noindent
where $\nu_j$ is the density of states in the $j$'th HD layer and $\kappa$ is the background dielectric function of a thin film on SiC ($\kappa \approx (10+1)/2 = 5.5$). Because of the large difference in carrier density between layers near and far from the substrate, allowed $q$ values are much smaller than $q^{\rm{MEG}}_{\rm{TF}}$ and the screened interlayer interaction reduces to $(2 \pi e^2)/(\kappa  q^{\rm{MEG}}_{\rm{TF}})$.

The remaining integrals in Equation~\ref{eq:eratediffappend} contain a logarithmic divergence at zero wavevector. This is removed using a cutoff of $\omega/v_{\rm F}$ which reflects the zero value of ${\rm Im}[\chi(q ,\omega)]$ above the intraband particle-hole continuum.

Finally, we find that the linearized interlayer energy transfer rate per area between a pair of graphene layers in a MEG system (where one is HD and the other LD) is: 
\begin{equation}
\begin{split}\label{eq:DegenErateAppend}
\frac{1}{L^2} \cdot \cQ^{\rm{el}}_{\rm{HD, LD}} = & \gamma (T_{\rm{LD}}-T_{\rm{HD}}) \left( \frac{\nu_{\rm{LD}} \nu_{\rm{HD}}}{(\sum_{j}\nu_j)^2} \right) \\
& T_{\rm{HD}}^3 \ln \left( \frac{T_{\rm{F, LD}}}{T_{\rm{HD}}} \right),
\end{split}
\end{equation}
\noindent
where $\gamma = (8 \pi^2 / 30)(k_{\rm B}^4/\hbar^3 v_{\rm F}^2)$. The density of states in the HD layer (LD layer) is denoted by $\nu_{\rm{HD}}$ ($\nu_{\rm{LD}}$). The sum over index $j$ runs over all HD layers in the MEG system, accounting for the approximation that the dominant screening effect originates from the majority fraction of carriers in the HD layers (see the main text). $T_{\rm{F, LD}} = E_{\rm{F, LD}}/k_{\rm B}$ is the Fermi temperature of the LD layer. Equation~\ref{eq:DegenErateAppend} becomes exact in the degenerate limit  $T \ll T_{\rm{F,LD}}, T_{\rm{F,HD}}$.

%% Supplementary Note 3
\newpage
\subsection{Supplementary Note 3 \\ Interlayer energy transfer to leading order in $T$}
Here, we derive in detail the leading order in temperature formula for the interlayer Coulombic energy transfer rate between one lightly doped (LD) layer of graphene and one highly doped (HD) layer of graphene. This follows similar steps as in the previous section, but without the assumption of infinitesimal temperature separation between the LD and HD layers. As described in the main text, the interlayer energy transfer rate per area between an LD and an HD layer is: 
\begin{equation}
\begin{split}\label{eq:appendsec3eq1}
\frac{\cQ^{\rm{el}}}{L^2} = & \frac{\hbar}{4 \pi^3}\int_{-\infty}^{\infty}\omega \, d\omega \int d\vec{q} \, \vert v^{\rm{sc}}_{\rm{LD,HD}}\vert^2 \left[n_{\rm B}(T_{\rm{LD}})-n_{\rm B}(T_{\rm{HD}}) \right] \\
& \times \quad {\rm Im}[\chi_{\rm{LD}}(q,\omega,T_{\rm{LD}})] {\rm Im}[\chi_{\rm{HD}}(q,\omega,T_{\rm{HD}})],
\end{split}
\end{equation}
\noindent
where $n_{\rm B}(T)$ is a Bose distribution function and $v^{\rm{sc}}_{\rm{LD,HD}} = v_q/\epsilon^{\rm{RPA}}(q,\omega,T_{\rm{LD}},T_{\rm{HD}})$ is the screened interaction between electrons in opposing layers within the random phase approximation (RPA). Given that the separation of the two layers is a distance $d$, the RPA dielectric function is given by: 
\begin{equation}
\begin{split}\label{eq:appendsec3eq2}
\epsilon^{\rm{RPA}}(q,\omega,T_{\rm{LD}},T_{\rm{HD}})= & (1-v_q \chi_{\rm{LD}}(q,\omega,T_{\rm{LD}}))(1-v_q \chi_{\rm{HD}}(q,\omega,T_{\rm{HD}})) \\
&  \quad - \, v^2_q e^{-2 q d}\chi_{\rm{LD}}(q,\omega,T_{\rm{LD}}) \chi_{\rm{HD}}(q,\omega,T_{\rm{HD}}),
\end{split}
\end{equation}
\noindent
where $v_q = 2 \pi e^2/\kappa q$ and $\chi_i(q,\omega,T)$ is the temperature dependent non-interacting density-response function of the $i$'th layer of graphene. We next assume the temperature of the electrons in the HD layer is pinned to the lattice temperature, and approximate this as zero relative to the high temperature in the LD layer, i.e. $T_{\rm{HD}} = T_{\rm{L}} \rightarrow 0$. Relabeling $T_{\rm{LD}}$ as $T$ we have: 
\begin{equation}
\begin{split}\label{eq:appendsec3eq3}
\frac{\cQ^{\rm{el}}}{L^2} = & \frac{\hbar}{4 \pi^3}\int_{-\infty}^{\infty}\omega \, d\omega \int d\vec{q} \, \vert \frac{v_q e^{-q d}}{\epsilon^{\rm{RPA}}(q,\omega,T)}\vert^2 \, n_{\rm B}(T) \\
& \times \quad {\rm Im}[\chi_{\rm{LD}}(q,\omega,T)] {\rm Im}[\chi_{\rm{HD}}(q,\omega)].
\end{split}
\end{equation}
\noindent
In the degenerate limit, $k_{\rm B} T << E_{\rm{F,LD}}$, the Bose distribution function limits the important frequencies to approximately $\omega \le E_{\rm{F,LD}}/\hbar$, and reveals that it is the leading order in $\omega$ which will yield the leading order in $T$. With this motivation we make use of the low frequency limit, ${\rm Im}[\chi_i(q,\omega \rightarrow 0)] = -\nu_i \omega / (v_{\rm F} q)$, where $\nu_i = 2 E_{{\rm F},i}/(\pi \hbar^2 v_{\rm F}^2)$ is the density of states at the Fermi energy. Similarly, the dielectric function can be reduced to: 
\begin{equation}
\epsilon^{\rm{RPA}}(q,\omega,T) \rightarrow (1+\frac{q^{\rm{TF}}_{\rm{LD}}}{q})(1+\frac{q^{\rm{TF}}_{\rm{HD}}}{q})-e^{-2 q d}\frac{q^{\rm{TF}}_{\rm{LD}}q^{\rm{TF}}_{\rm{HD}}}{q^2},
\end{equation}
\noindent
where the Thomas-Fermi wavevector of the $i$'th layer is defined as $q^{\rm{TF}}_{i} = q v_q \nu_i$. Making use of the fact that in our MEG samples $k_{\rm{F,LD}}d << 1$ and $q^{\rm{TF}}_{\rm{HD}}/k_{\rm{F,LD}} >> 1$, we simplify the interlayer energy transfer rate per area to: 
\begin{equation}
\frac{\cQ^{\rm{el}}}{L^2} =\frac{E^4_{\rm{F,LD}} \nu_{\rm{LD}}}{2 \pi^2 v_{\rm F}^2 \hbar^3 \nu_{\rm{HD}}}\int_0^{\infty}\Omega d\Omega \int_{0}^{\infty} Q dQ \left(\frac{1}{e^{\Omega/t}-1}\right)\left( \frac{\Omega}{Q}\right)^2,
\end{equation}
\noindent
where we have introduced the dimensionless wavevector, $Q=q/k_{\rm{F,LD}}$, the dimensionless frequency, $\Omega = \hbar \omega/E_{\rm{F,LD}}$, and the dimensionless temperature, $t=k_{\rm B} T/E_{\rm{F,LD}}$. The wavevector integration here diverges logarithmically. To remedy this, we identify that ${\rm Im}[\chi_{\rm{LD}}(q,\omega,0)]$ vanishes for $\omega > v_{\rm F} q $, which cuts off the wavevector integration for $Q<\Omega$. Carrying out the remaining integrals, we obtain the leading order in temperature energy loss rate per area of the LD electrons: 
\begin{equation}
\frac{\cQ^{\rm{el}}}{L^2} =-\frac{E^4_{\rm{F,LD}} \pi^2 \nu_{\rm{LD}}}{15 v_{\rm F}^2 \hbar^3 \nu_{\rm{HD}}}t^4 \ln{t}.
\end{equation}
\noindent

%% Supplementary Note 4
\newpage
\subsection{Supplementary Note 4 \\ Interlayer energy transfer with no LD-LD layer coupling}
Here, we illustrate the distance dependence of the interlayer Coulombic energy transfer rate by calculating the temperature dynamics of an individual LD layer coupled to the HD layers in MEG systems, when we neglect energy transfer between pairs of LD layers. Although Coulomb coupling between LD layers is very strong \cite{SoljacicSI}, by temporarily forcing $\cQ^{\rm{el}}_{\rm{LD, LD^{\prime}}} \rightarrow 0$ we can gain insight into the distance dependence of the interlayer energy transfer rate. The temperature dynamics of each individual LD layer coupled to the HD layers are then independent and obey: 
\be
\partial_t T_{\rm{LD}} = \left(\sum_{j \in {\rm HD}} \cQ^{\rm{el}}_{{\rm LD}, j}(T_{\rm{HD}}=T_{\rm{L}},T_{\rm{LD}},d_{j , {\rm LD}})\right)/\cC_{\rm{LD}}.   \label{eq:Tlight}
\ee
\noindent
If we also approximate the heat capacity in the LD layers
by the neutral graphene formula $\cC_{\rm{LD}}=18 \zeta(3) T_{\rm{LD}}^2/(\pi v_{\rm F}^2)$, we obtain the results shown in Supplementary Figure~\ref{fig:figS3}, where we have used this formula to evaluate $T_{\rm{LD}}(t)$ for
several different values of $d_{\rm{HD, LD}}$.  All of these curves are calculated for
the typical carrier density of the LD layers in MEG, measured in experiment, of $n_{\rm{LD}} = 10^{10} {\rm cm}^{-2}$. 
The slowest ($d_{\rm{HD, LD}}=60$ layers) and fastest ($d_{\rm{HD, LD}}=1$ layer) temperature
relaxation curves provide upper and lower bounds, respectively, on the true relaxation
time of \emph{all} LD layers when interlayer energy transfer between pairs of LD layers is no longer neglected, i.e. $\cQ^{\rm{el}}_{\rm{LD, LD^{\prime}}} \neq 0$. 

%% Supplementary Note 5
\newpage
\subsection{Supplementary Note 5 \\ Acoustic phonon pinning of $T_{\rm{HD}}$ to $T_{\rm L}$}
Here, we present some simple calculations in support of our assumption that acoustic phonon cooling is capable of pinning the electronic temperature in the HD layers at the lattice temperature while these layers act as a heat sink for energy dissipation from the remaining hot LD layers. Heuristically, we also note that previous ultrafast optical spectroscopy experiments \cite{SunPSSC2011SI} have observed thermal relaxation times in the HD layers of MEG of $\sim 10$ ps, \emph{much} faster than the relaxation times in the LD layers of $\sim 100-500$ ps that we report here.

Acoustic phonon cooling has previously been investigated in the context of hot-carrier cooling in disorder-free monolayer graphene \cite{BistritzerPRL2009SI,TsePRB2009SI}. In these single layer systems, as a result of the relatively large optical phonon energy in graphene ($\hbar\omega_{\rm{op}} \approx 200$ meV \cite{deHeerAPL2008SI}), acoustic phonons serve as the primary intrinsic cooling mechanism over a large temperature window extending up towards $T \sim 250$ K \cite{BistritzerPRL2009SI}. To estimate the ability of acoustic phonon cooling to take away the electronic energy that is Coulomb-transfered from LD to HD layers, we use Equation 14 in Bistritzer and MacDonald \cite{BistritzerPRL2009SI} to calculate the total energy transfer rate to the the lattice. Using the carrier density profile of the four most highly doped layers measured in experiment ($E_{{\rm F},1-4} = 360, 218, 140,$ and $93$ meV), we find that $\cQ^{\rm{ph}}_{\rm{HD}}= 9.09 (T_{\rm{HD}}-T_{\rm L})$ W cm$^{-2}$ can be transferred to the lattice via carrier-phonon scattering in the HD layers. We can calculate the ratio of $\cQ^{\rm{ph}}$ to the interlayer energy transfer rate $\cQ^{\rm{el}}$ from the $N$ LD layers to these four HD layers using Equation 13 in the main text. Supplementary Figure~\ref{fig:figS4} suggests that for both $N=30$ and $N=60$ layer MEG systems, the acoustic phonon cooling power is sufficient to keep the HD layers pinned to the lattice temperature while absorbing energy from the distant LD layers.

%% Supplementary Note 6
\newpage
\subsection{Supplementary Note 6 \\ Disorder-assisted electron-phonon (supercollision) cooling}
Here, we present an application of the recently proposed disorder-assisted electron-phonon (supercollision) cooling mechanism \cite{LevitovPRL2012SI} to MEG. We investigate the qualitative and quantitative differences between disorder-assisted electron-phonon cooling in HD and LD graphene and we clearly demonstrate that this cooling mechanism cannot alone explain electronic cooling in high quality MEG. 

The electron temperature dynamics of a single graphene layer in the framework of the disorder-assisted electron-phonon cooling mechanism is given by: 
\begin{equation}
\partial_{t} T = \cQ^{\rm{sc}}/\cC,    \label{eq:dTsc}
\end{equation}
where $\cQ^{\rm{sc}} = \partial_{t} \cE$ is the electronic cooling rate due to supercollisions, $\cC = \partial_{T} \cE$ is the electronic heat capacity and $\cE$ is the electronic energy density. The electronic cooling rate due to supercollisions in the degenerate limit when $k_{\rm{B}}T \ll E_{\rm{F}}$ is given by \cite{LevitovPRL2012SI}: 
\begin{equation}
\cQ^{\rm{sc}} = -A(T^{3} - T_{\rm{L}}^{3}),    \label{eq:QscDeg}
\end{equation}
with a rate coefficient $A = 9.62g^{2}\nu^{2}(E_{\rm{F}})k_{\rm{B}}^{3} / (\hbar k_{\rm{F}}l)$, where $g = D / \sqrt{2\rho v_{\rm{s}}^{2}}$ is the electron-phonon coupling constant and $\nu(E_{\rm{F}}) = E_{\rm{F}} / (2\pi\hbar^{2}v_{\rm{F}}^{2})$ is the density of states at the Fermi level per one spin and one valley flavor. The rest of the parameters are the Fermi velocity $v_{\rm{F}}$, the sound velocity $v_{\rm{s}}$, the deformation potential $D$, the mass density $\rho$ and the disorder mean free path $l$. The electronic cooling rate in the non-degenerate limit when $k_{\rm{B}}T \gg E_{\rm{F}}$ is given by \cite{HakonenNanoLett2014SI}: 
\begin{equation}
\cQ^{\rm{sc}} = -B(T^{5} - T_{\rm{L}}^{5}),    \label{eq:QscNonDeg}
\end{equation}
with a rate coefficient $B = ((4k_{\rm{B}}^{2}) / (E_{\rm{F}}^{2})) A$. Because both the supercollision cooling rate and the electronic heat capacity have different functional dependence on the electron temperature at high and at low doping densities, we need to consider the two cases separately. 

For HD graphene, the heat capacity is given by $\cC = \alpha T = ((2\pi E_{\rm{F}}k_{\rm{B}}^{2}) / (3\hbar^{2}v_{\rm{F}}^{2}))T$. By substituting this expression and Equation~\ref{eq:QscDeg} in Equation~\ref{eq:dTsc}, we obtain \cite{GrahamNatPhys2013SI}: 
\begin{equation}
\partial_{t} T = -\frac{A}{\alpha}\frac{T^{3} - T_{\rm{L}}^{3}}{T}.		\label{eq:dTscHD}
\end{equation}
The electronic cooling timescale is governed by the rate coefficient $A / \alpha \propto E_{\rm{F}} / k_{\rm{F}}l \propto 1 / l$. 
In the high and low electron temperature limits, Equation~\ref{eq:dTscHD} reduces to: 
\begin{equation}
\partial_{t} T = -\frac{A}{\alpha}T^{2}	\qquad	for	\qquad	T \gg T_{\rm{L}},		\label{eq:dTscHDhigh}
\end{equation}
\begin{equation}
\partial_{t} T = -\frac{3A}{\alpha}T_{\rm{L}}(T - T_{\rm{L}})	\qquad	for	\qquad	T \approx T_{\rm{L}},		\label{eq:dTscHDlow}
\end{equation}
with solutions given by: 
\begin{equation}
T(t) = \frac{T_{0}}{1 + \frac{A}{\alpha}T_{0}t}	\qquad	for	\qquad	T \gg T_{\rm{L}},		\label{eq:TscHDhigh}
\end{equation}
\begin{equation}
T(t) = T_{\rm{L}} + (T_{0} - T_{\rm{L}})\exp\left(-\frac{3A}{\alpha}T_{\rm{L}}t\right)	\qquad	for	\qquad	T \approx T_{\rm{L}}.		\label{eq:TscHDlow}
\end{equation}

For LD graphene, the heat capacity is approximated by $\cC = \beta T^{2} = ((18\zeta(3) k_{\rm{B}}^{3}) / (\pi\hbar^{2}v_{\rm{F}}^{2}))T^{2}$. By substituting this expression and Equation~\ref{eq:QscNonDeg} in Equation~\ref{eq:dTsc}, we obtain: 
\begin{equation}
\partial_{t} T = -\frac{B}{\beta}\frac{T^{5} - T_{\rm{L}}^{5}}{T^{2}}.		\label{eq:dTscLD}
\end{equation}
The electronic cooling timescale is governed by the rate coefficient $B / \beta \propto 1 / k_{\rm{F}}l \propto 1 / E_{\rm{F}}l$. 
In the high and low electron temperature limits, Equation~\ref{eq:dTscLD} reduces to: 
\begin{equation}
\partial_{t} T = -\frac{B}{\beta}T^{3}	\qquad	for	\qquad	T \gg T_{\rm{L}},		\label{eq:dTscLDhigh}
\end{equation}
\begin{equation}
\partial_{t} T = -\frac{5B}{\beta}T_{\rm{L}}^{2}(T - T_{\rm{L}})	\qquad	for	\qquad	T \approx T_{\rm{L}},		\label{eq:dTscLDlow}
\end{equation}
with solutions given by: 
\begin{equation}
T(t) = \frac{T_{0}}{\sqrt{1 + 2\frac{B}{\beta}T_{0}^{2}t}}	\qquad	for	\qquad	T \gg T_{\rm{L}},		\label{eq:TscLDhigh}
\end{equation}
\begin{equation}
T(t) = T_{\rm{L}} + (T_{0} - T_{\rm{L}})\exp\left(-\frac{5B}{\beta}T_{\rm{L}}^{2}t\right)	\qquad	for	\qquad	T \approx T_{\rm{L}}.		\label{eq:TscLDlow}
\end{equation}

We observe that the different functional form of the cooling rate and the heat capacity at high and at low doping densities results in qualitatively different electron temperature relaxation dynamics. In the low electron temperature limit, in particular, the temperature dynamics follows an exponential form with a lattice temperature dependent characteristic time $\tau_{\rm{HD}} = ((3A/\alpha)T_{\rm{L}})^{-1} \propto l / T_{\rm{L}}$ for HD graphene, and with a lattice temperature dependent characteristic time $\tau_{\rm{LD}} = ((5B/\beta)T_{\rm{L}}^{2})^{-1} \propto E_{\rm{F}}l / T_{\rm{L}}^{2}$ for LD graphene. 

To obtain quantitative estimates of the full electron temperature dynamics predicted by the disorder-assisted electron-phonon cooling mechanism, we solve Equation~\ref{eq:dTscHD} and Equation~\ref{eq:dTscLD} numerically. In all calculations, we use $v_{\rm{F}} = 1 \times 10^6$ m s$^{-1}$, $v_{\rm{s}} = 2.1 \times 10^4$ m s$^{-1}$, $\rho = 7.6 \times 10^{-7}$ kg m$^{-2}$ and $D = 20$ meV. Supplementary Figure~\ref{fig:figS5} and Supplementary Figure~\ref{fig:figS6} show the calculated electron temperature dynamics for HD graphene with $E_{\rm{F}} = 100$ meV and for LD graphene with $E_{\rm{F}} = 10$ meV, respectively, at $T_{\rm{L}} = 10$ K for variable disorder mean free path $l$. The large uncertainty in the value of the disorder length scale leads to a very large spread in the calculated electron temperature dynamics. 

The experimental measurement of the precise value of the disorder mean free path between supercollisions is a challenging task and reliable estimates are lacking in the literature. One reliable way to characterize the degree of disorder which can potentially be related to supercollision cooling is directly from the width of the Dirac cone near the Dirac point in the graphene band structure as measured in high-resolution angle-resolved photoemission spectroscopy (ARPES). We use such recent ARPES measurements to estimate the disorder mean free path value appropriate for our MEG samples \cite{deHeerNature2014SI}. In Ref. \citenum{deHeerNature2014SI} (Figure S1), correlation lengths of $\sim 1-3$ nm in exfoliated graphene and correlation lengths exceeding $\sim 50$ nm (limited {\em only} by the instrument resolution, but are expected to be even longer) in C-face MEG have been reported. Supplementary Figure~\ref{fig:figS7} shows the calculated electron temperature dynamics for HD graphene with $E_{\rm{F}} = 100$ meV and disorder mean free path $l = 5$ nm for variable $T_{\rm{L}}$. We note that these calculations are consistent with recent experimental results based on photocurrent measurements on HD chemical-vapor-deposited (CVD) graphene \cite{GrahamNatPhys2013SI}. Supplementary Figure~\ref{fig:figS8} shows the calculated electron temperature dynamics for LD graphene with $E_{\rm{F}} = 10$ meV and disorder mean free path $l = 50$ nm for variable $T_{\rm{L}}$. 

We also present the electronic cooling times in the low electron temperature limit predicted by the disorder-assisted electron-phonon cooling mechanism that are more straightforward to compare to the experiment. Supplementary Figure~\ref{fig:figS9} and Supplementary Figure~\ref{fig:figS10} show the calculated electronic cooling times $\tau_{\rm{HD}}$ for HD graphene with $E_{\rm{F}} = 100$ meV and $\tau_{\rm{LD}}$ for LD graphene with $E_{\rm{F}} = 10$ meV, respectively, as a function of $T_{\rm{L}}$ for variable disorder mean free path $l$. We observe that the predictions of the supercollision cooling model applied to the HD layers of MEG are inconsistent both in magnitude and lattice temperature dependence with the ultrafast time-resolved THz spectroscopy experiments. We also observe that the model predictions for the LD layers of MEG can roughly capture the order of magnitude and the lattice temperature dependence for some disorder mean free path, but not the layer number dependence. Because the quality of our MEG samples is expected to be even higher than we have conservatively assumed in these calculations, we conclude that the disorder-assisted electron-phonon (supercollision) cooling can provide a parallel cooling channel, but it is not dominant in MEG. 

%%%%%%%%%%%%%%%%%%%%%%%%%%%%%%%%%%%%%%%%%%%%%%%%%%%%%%%%%%%%%%%%%%%%%
%% Supplementary Figures
%%%%%%%%%%%%%%%%%%%%%%%%%%%%%%%%%%%%%%%%%%%%%%%%%%%%%%%%%%%%%%%%%%%%%
%\section{Supplementary Figures}

% Supplementary Figure 1

% Supplementary Figure 2

% Supplementary Figure 3
\begin{figure}
\begin{center}
%   \begin{tabular}{@{}cc@{}}
%\includegraphics[height=2.1in]{Figure_S3.png}
\includegraphics[width=4.0in]{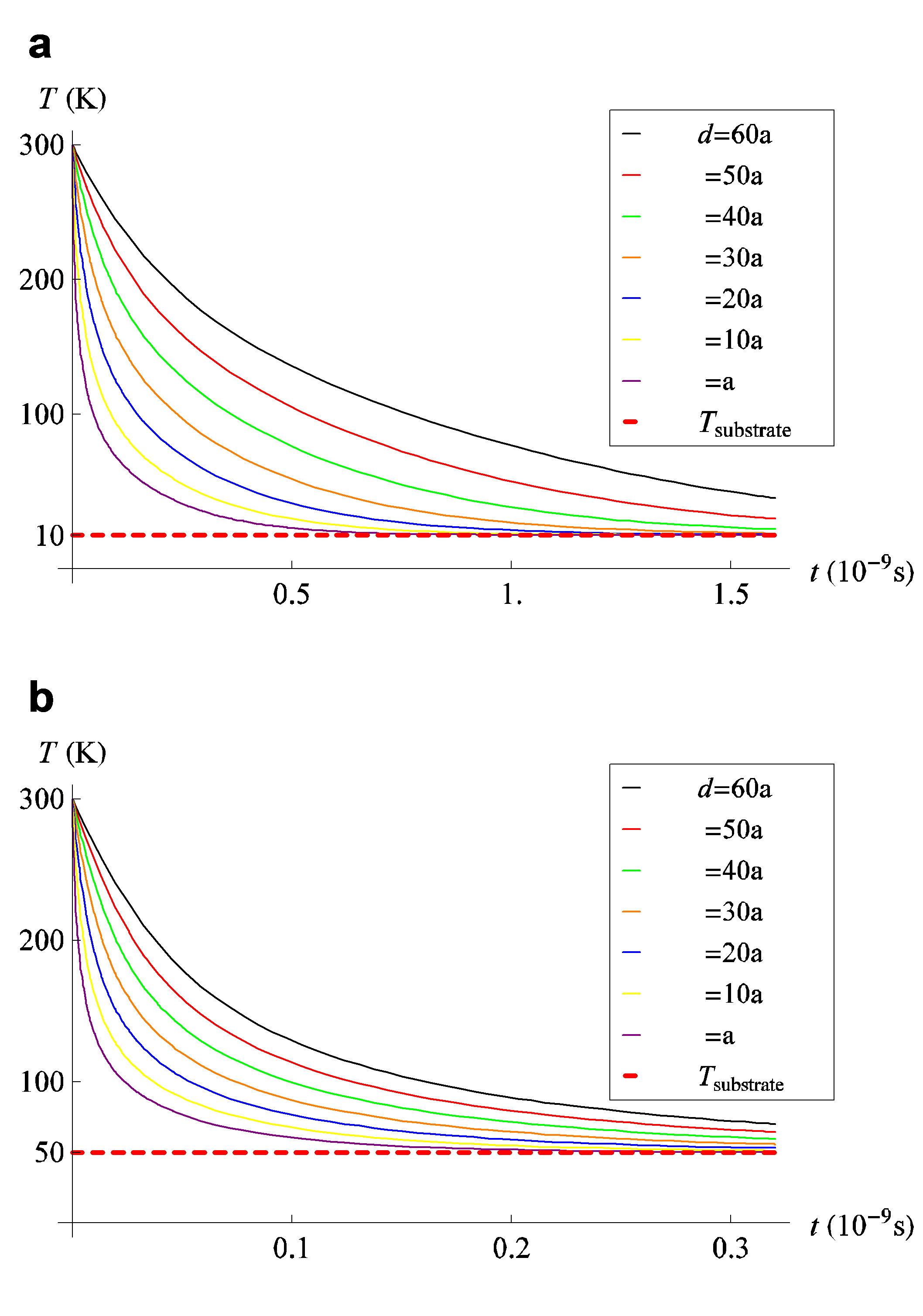}
% \end{tabular}
\caption{{\bf Interlayer energy transfer with no LD-LD layer coupling.} 
Temperature dynamics when interlayer energy transfer between LD layers ($n_{\rm{LD}}=10^{10}$cm$^{-2}$) is ignored. The LD layer temperature $T_{\rm{LD}}(t)$ resulting from cooling via interlayer Coulombic energy transfer to the HD layers near the substrate ($n_{\rm{HD}} \gtrsim 10^{12}$cm$^{-2}$) at a constant lattice temperature $T_{\rm{L}}$. The distance between the particular LD layer and the HD layers is varied, $d_{\rm{HD, LD}}=60a, 50a, 40a, 30a, 20a, 10a, a$ (top to bottom) where $a=3.4$ Angstroms. Subfigure (a) shows results for lattice temperature $T_{\rm{L}}=10$ K and Subfigure (b) for $T_{\rm{L}}=50$ K.}
\label{fig:figS3}
\end{center}
\end{figure}

\clearpage
% Supplementary Figure 4
\begin{figure}
\begin{center}
\includegraphics[width=4.0in]{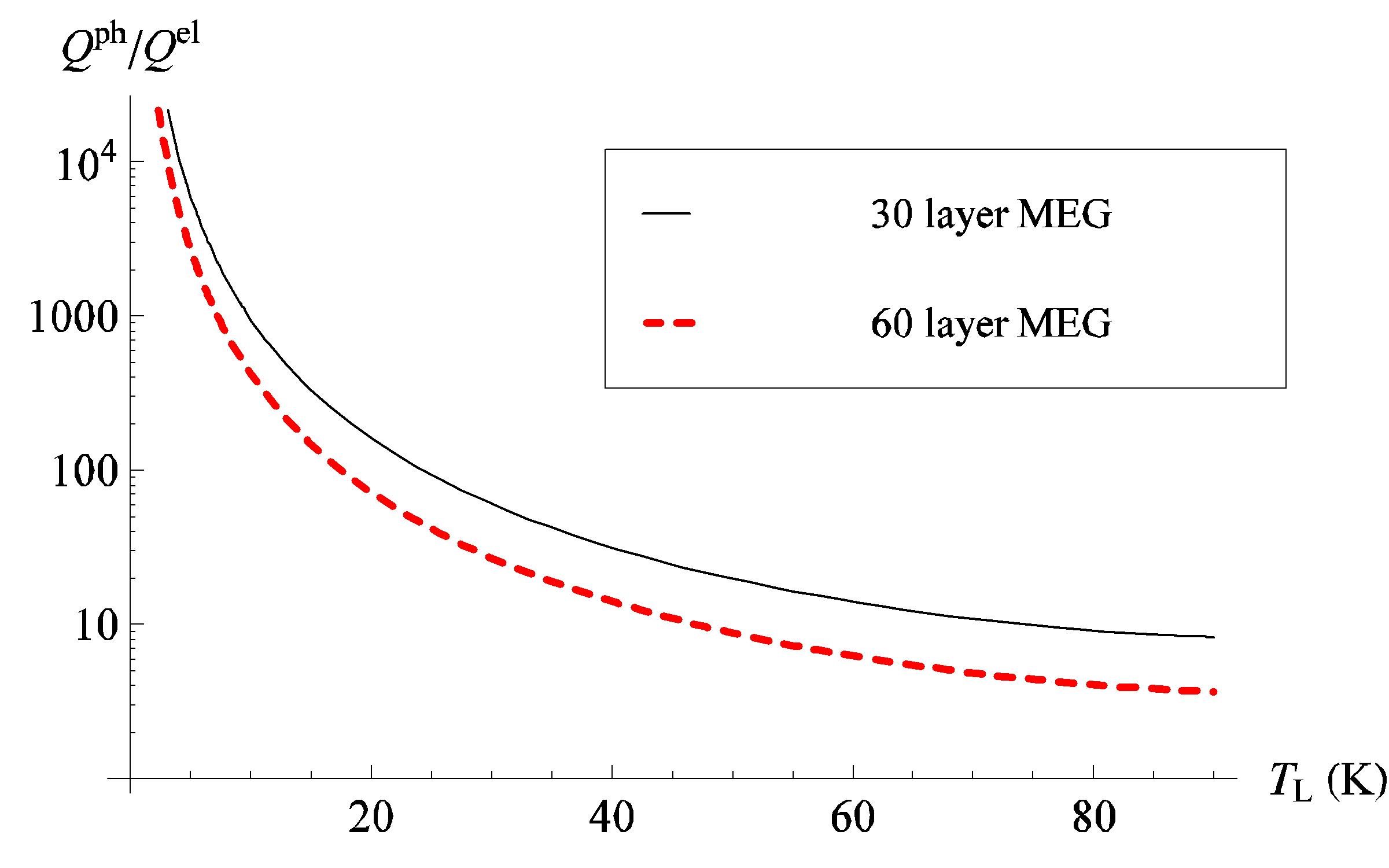}
\caption{{\bf Acoustic phonon pinning of $T_{\rm{HD}}$ to $T_{\rm L}$.} 
Ratio of the acoustic phonon cooling power $\cQ^{\rm{ph}}$ in the HD layers of MEG to the interlayer Coulombic energy transfer rate $\cQ^{\rm{el}}$ from the LD to the HD layers of MEG as a function of the lattice temperature $T_{\rm{L}}$.}
\label{fig:figS4}
\end{center}
\end{figure}

%\clearpage
% Supplementary Figure 5
\begin{figure}
\begin{center}
\includegraphics[width=4.0in]{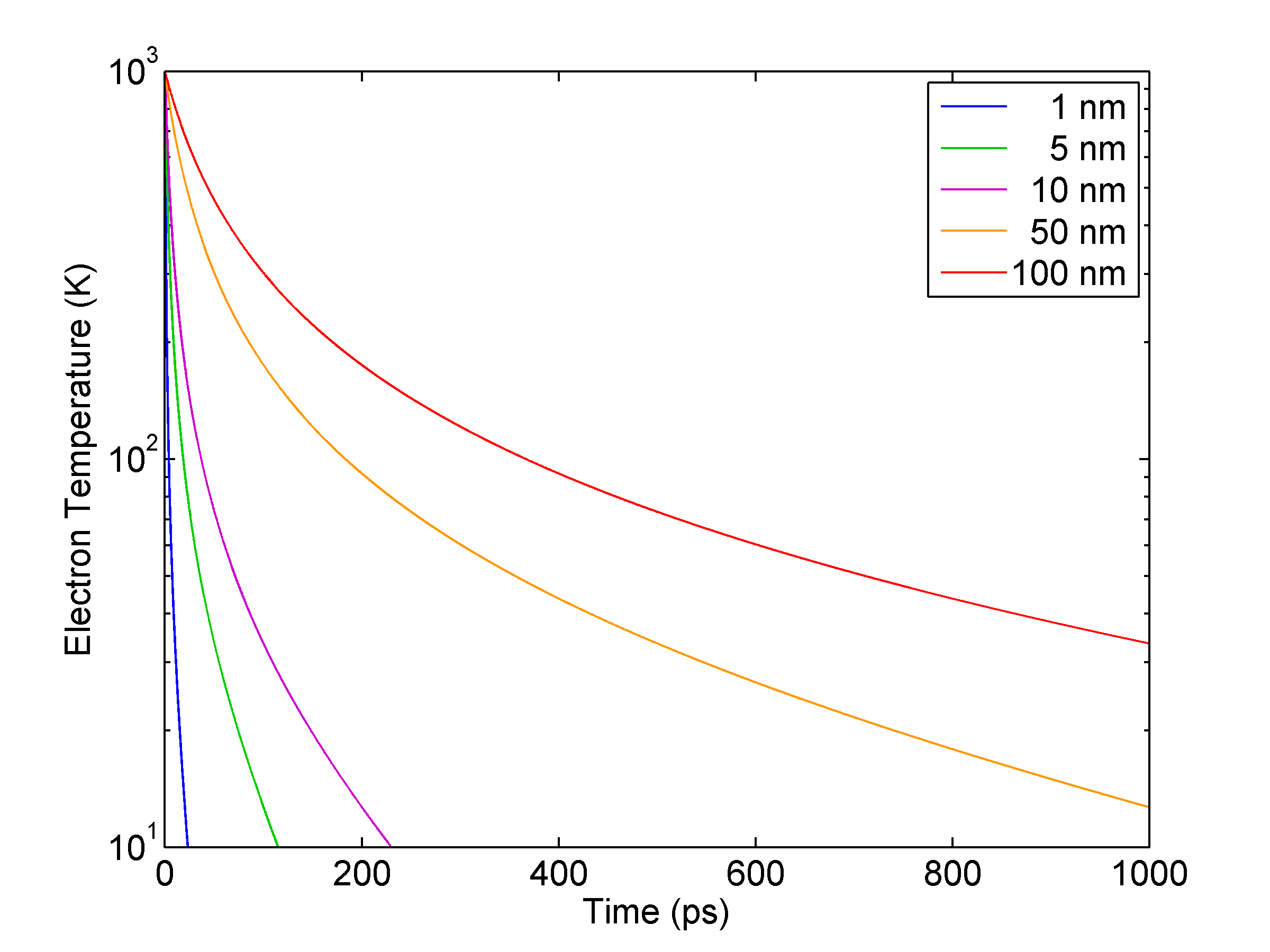}
\caption{{\bf Disorder-assisted electron-phonon (supercollision) cooling in HD graphene.} 
Electron temperature dynamics $T(t) - T_{\rm{L}}$ predicted by the disorder-assisted electron-phonon cooling mechanism for HD graphene with $E_{\rm{F}} = 100$ meV at $T_{\rm{L}} = 10$ K for variable disorder mean free path $l$.}
\label{fig:figS5}
\end{center}
\end{figure}

% Supplementary Figure 6
\begin{figure}
\begin{center}
\includegraphics[width=4.0in]{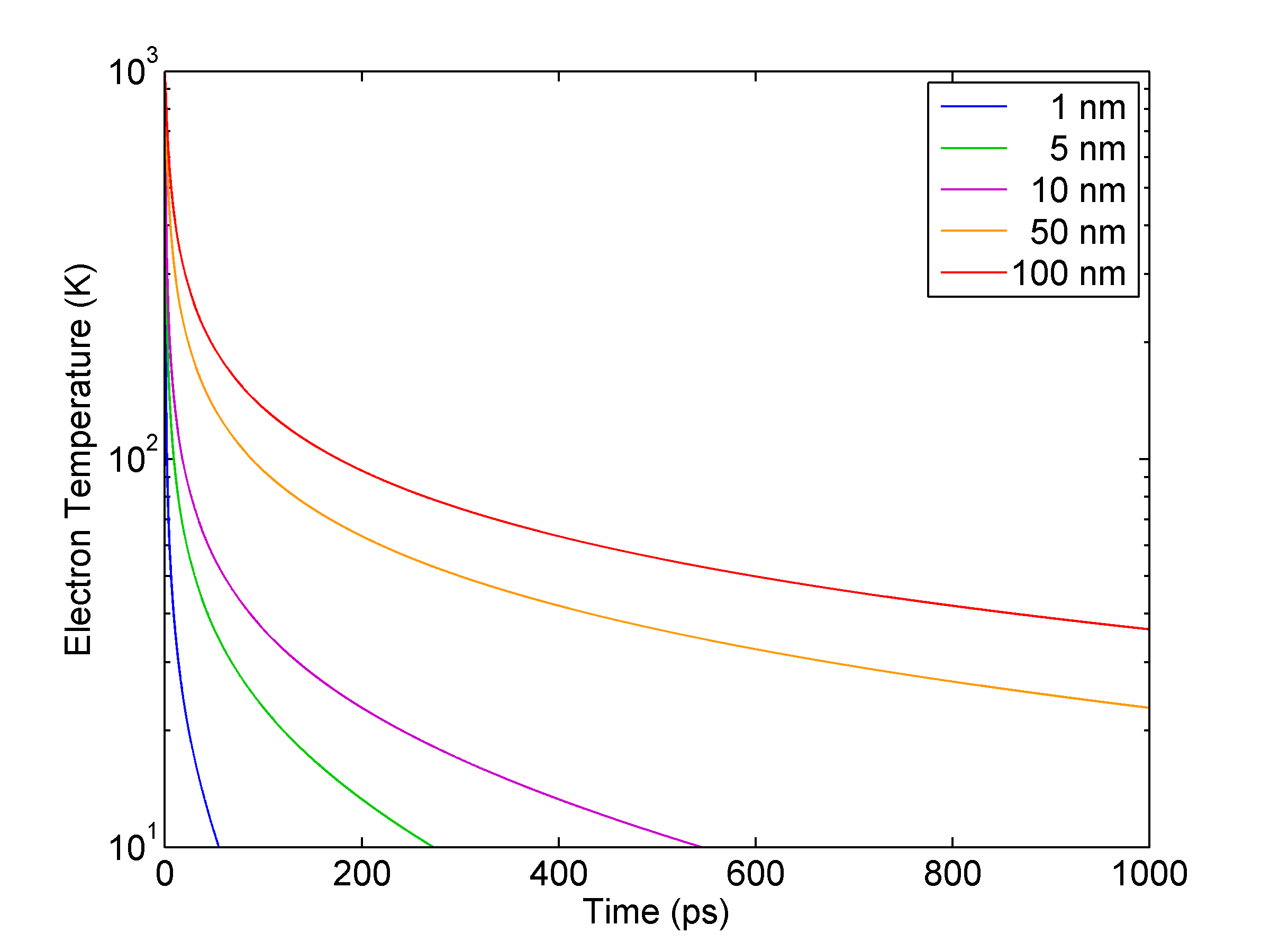}
\caption{{\bf Disorder-assisted electron-phonon (supercollision) cooling in LD graphene.} 
Electron temperature dynamics $T(t) - T_{\rm{L}}$ predicted by the disorder-assisted electron-phonon cooling mechanism for LD graphene with $E_{\rm{F}} = 10$ meV at $T_{\rm{L}} = 10$ K for variable disorder mean free path $l$.}
\label{fig:figS6}
\end{center}
\end{figure}

% Supplementary Figure 7
\begin{figure}
\begin{center}
\includegraphics[width=4.0in]{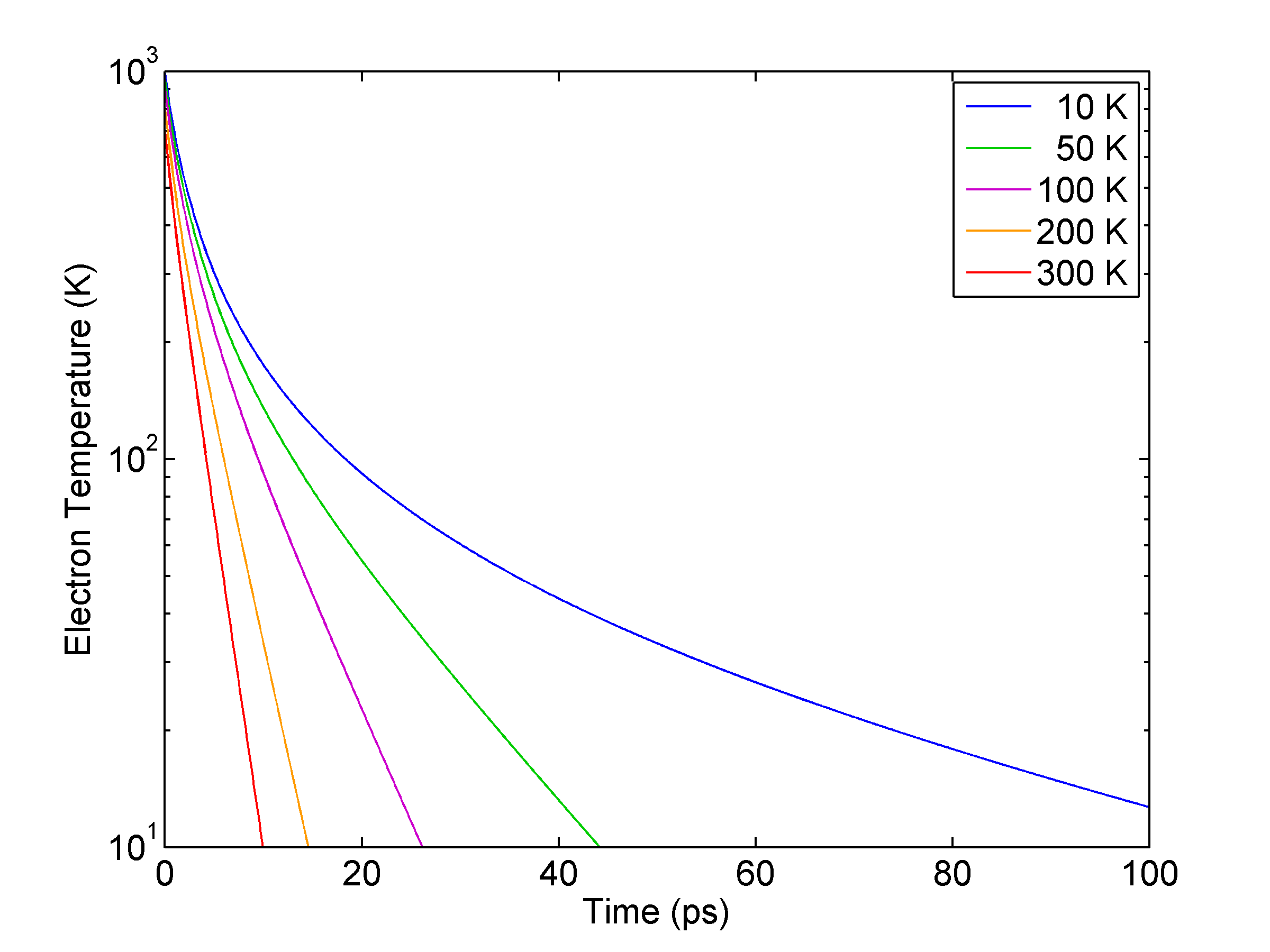}
\caption{{\bf Disorder-assisted electron-phonon (supercollision) cooling in HD graphene.} 
Electron temperature dynamics $T(t) - T_{\rm{L}}$ predicted by the disorder-assisted electron-phonon cooling mechanism for HD graphene with $E_{\rm{F}} = 100$ meV and disorder mean free path $l = 5$ nm for variable $T_{\rm{L}}$.}
\label{fig:figS7}
\end{center}
\end{figure}

% Supplementary Figure 8
\begin{figure}
\begin{center}
\includegraphics[width=4.0in]{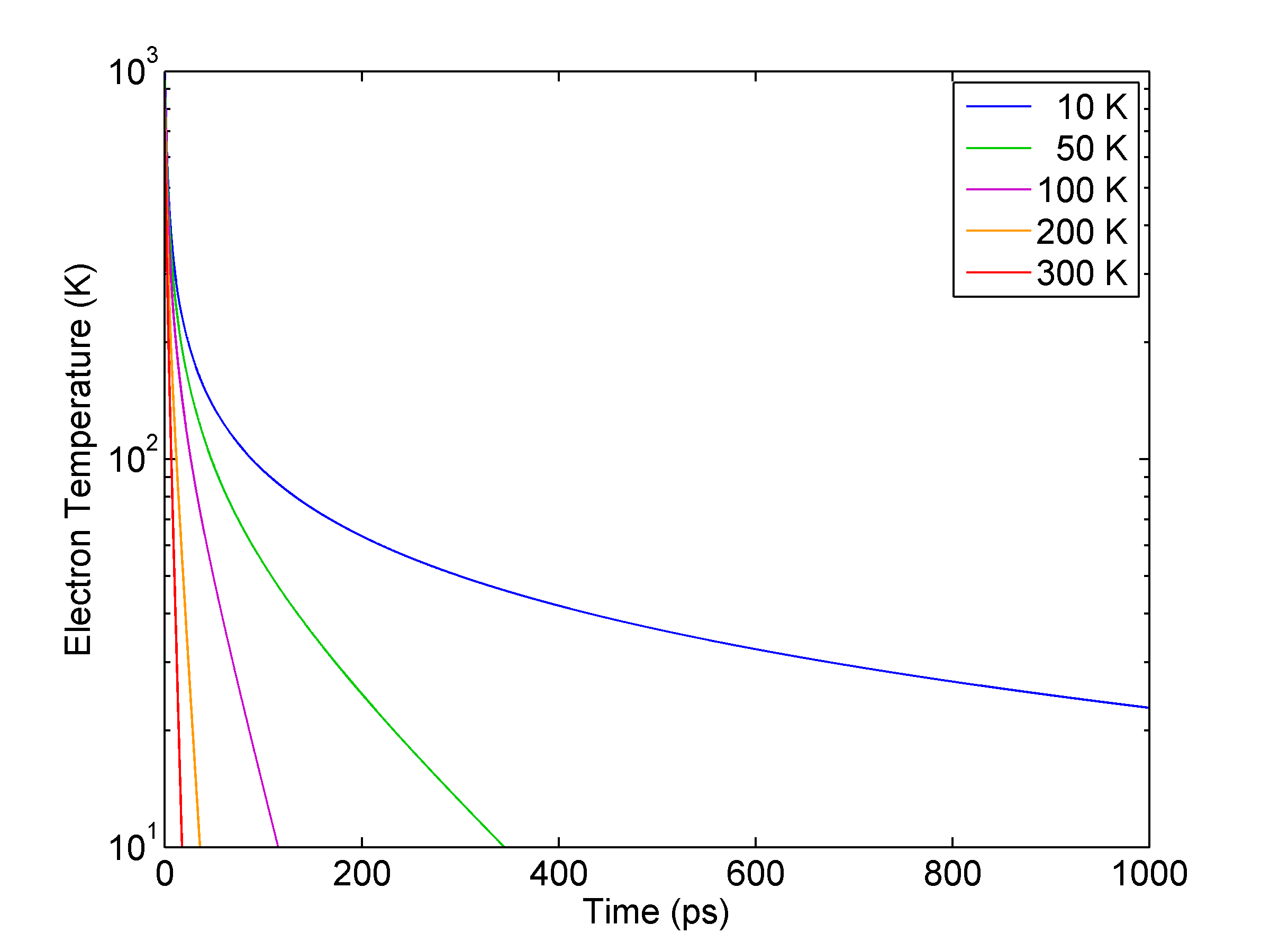}
\caption{{\bf Disorder-assisted electron-phonon (supercollision) cooling in LD graphene.} 
Electron temperature dynamics $T(t) - T_{\rm{L}}$ predicted by the disorder-assisted electron-phonon cooling mechanism for LD graphene with $E_{\rm{F}} = 10$ meV and disorder mean free path $l = 50$ nm for variable $T_{\rm{L}}$.}
\label{fig:figS8}
\end{center}
\end{figure}

% Supplementary Figure 9
\begin{figure}
\begin{center}
\includegraphics[width=4.0in]{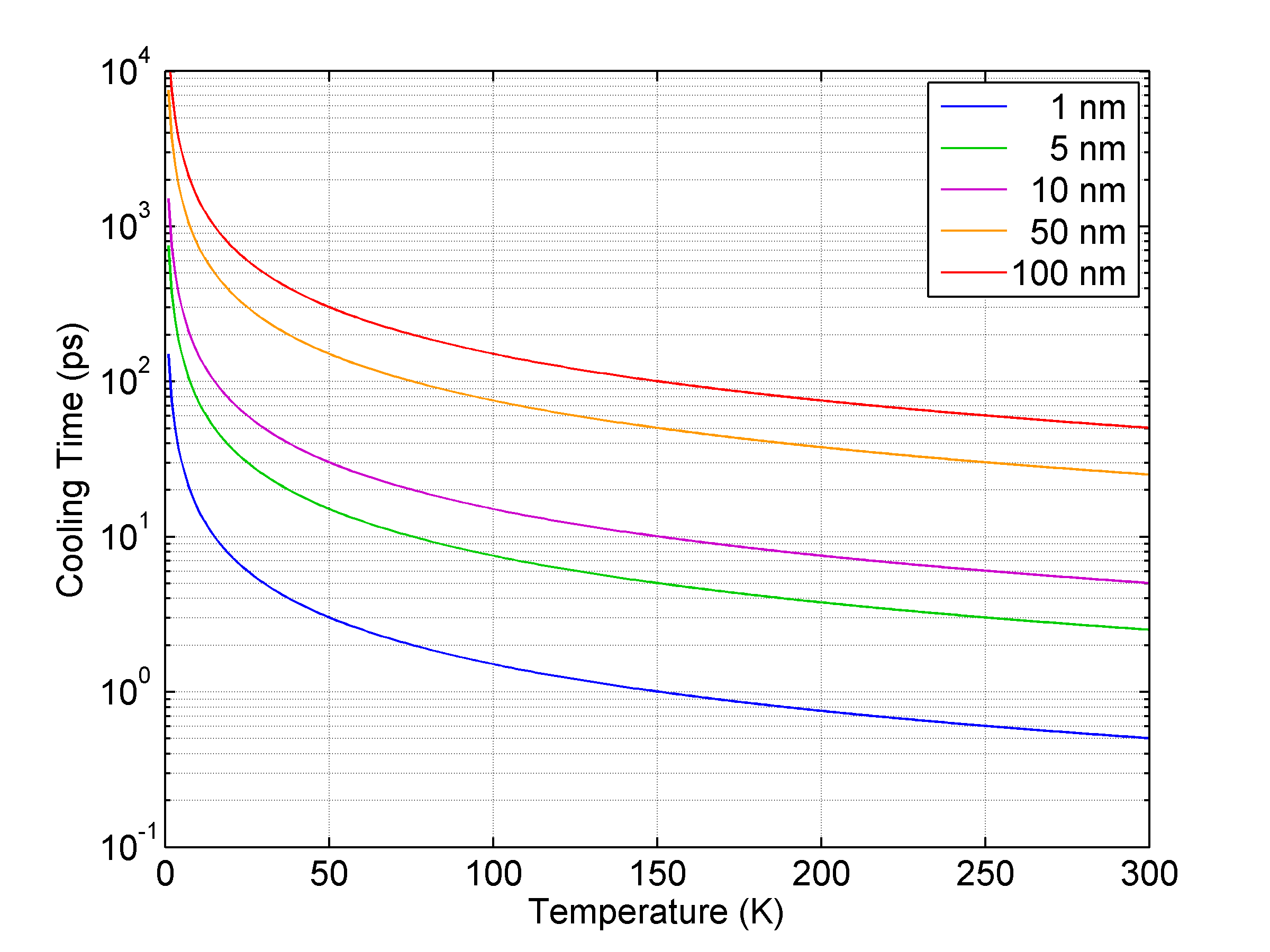}
\caption{{\bf Disorder-assisted electron-phonon (supercollision) cooling in HD graphene.} 
Electronic cooling time $\tau_{\rm{HD}}$ in the low electron temperature limit predicted by the disorder-assisted electron-phonon cooling mechanism for HD graphene with $E_{\rm{F}} = 100$ meV as a function of $T_{\rm{L}}$ for variable disorder mean free path $l$.}
\label{fig:figS9}
\end{center}
\end{figure}

% Supplementary Figure 10
\begin{figure}
\begin{center}
\includegraphics[width=4.0in]{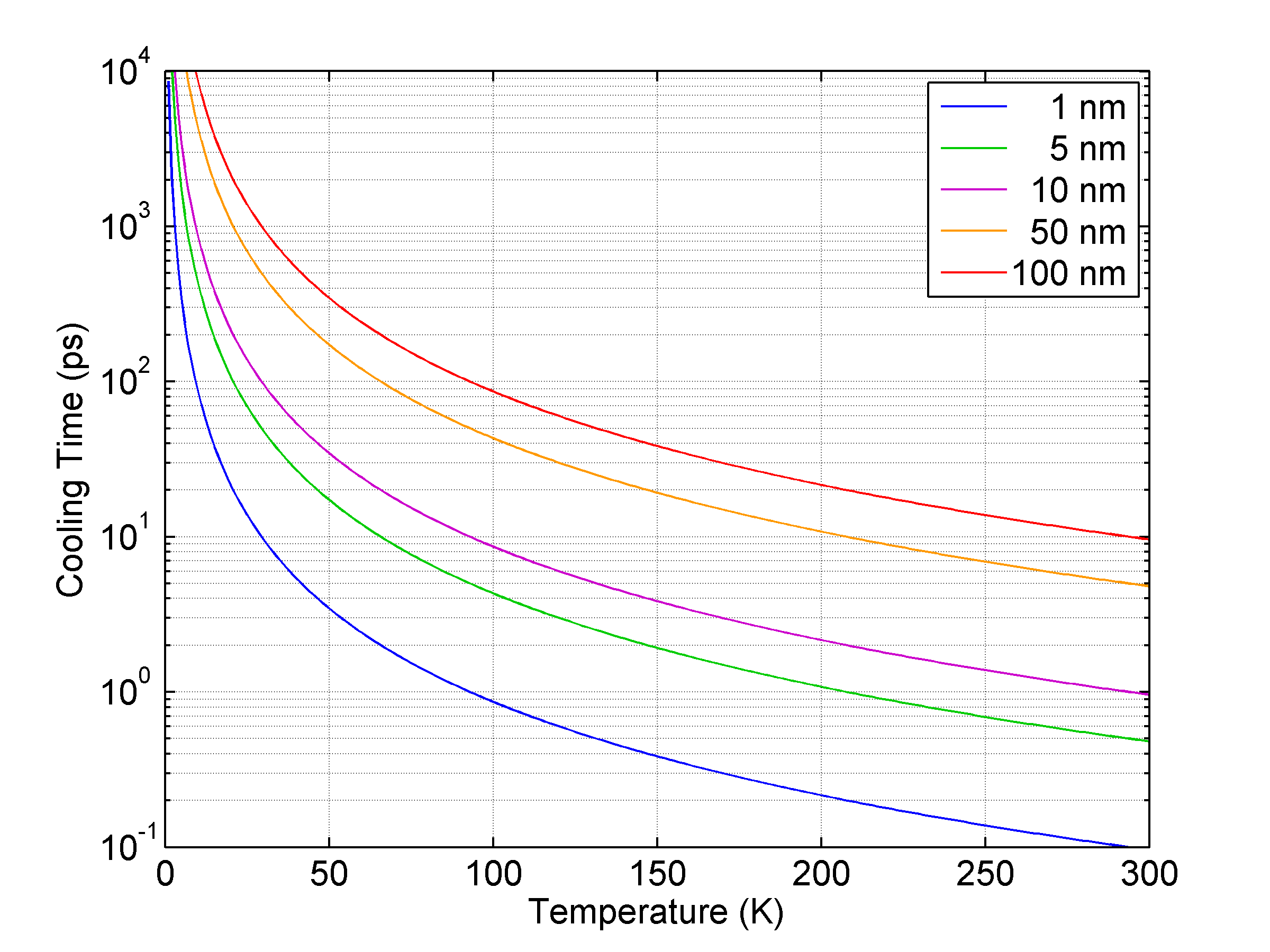}
\caption{{\bf Disorder-assisted electron-phonon (supercollision) cooling in LD graphene.} 
Electronic cooling time $\tau_{\rm{LD}}$ in the low electron temperature limit predicted by the disorder-assisted electron-phonon cooling mechanism for LD graphene with $E_{\rm{F}} = 10$ meV as a function of $T_{\rm{L}}$ for variable disorder mean free path $l$.}
\label{fig:figS10}
\end{center}
\end{figure}

%%%%%%%%%%%%%%%%%%%%%%%%%%%%%%%%%%%%%%%%%%%%%%%%%%%%%%%%%%%%%%%%%%%%%
%% Supplementary References
%%%%%%%%%%%%%%%%%%%%%%%%%%%%%%%%%%%%%%%%%%%%%%%%%%%%%%%%%%%%%%%%%%%%%
\clearpage
%\section{Supplementary References}

%%%%%%%%%%%%%%%%%%%%%%%%%%%%%%%%%%%%%%%%%%%%%%%%%%%%%%%%%%%%%%%%%%%%%
%% End
%%%%%%%%%%%%%%%%%%%%%%%%%%%%%%%%%%%%%%%%%%%%%%%%%%%%%%%%%%%%%%%%%%%%%
\end{document}